%% file: sedim-arXiv.tex
\documentclass[letterpaper]{article}

\usepackage{natbib}
\usepackage{rotating}
\usepackage{graphicx}
\usepackage{textcomp}
\usepackage{color}
\usepackage{Times}

\bibpunct{(}{)}{;}{a}{}{,}

\newcommand{\emm}[1]{\ensuremath{#1}}	
\newcommand{\emr}[1]{\emm{\mathrm{#1}}} 

\newcommand{\unit}[1]{\emm{\, \emr{#1}}}
\newcommand{\ala}{\mbox{\rlap{\hbox{\lower4pt\hbox{$\sim$}}}\hbox{$<$}}} 

\newcommand{\Msun}{\unit{M_\odot}}

\newcommand{\nht}{\ifmmode {{\rm NH}_3} \else {NH{\bas 3}} \fi}
\newcommand{\as}{\ifmmode {^{\scriptscriptstyle\prime\prime}}
	\else $^{\scriptscriptstyle\prime\prime}$\fi}

\newcommand{\psize}{\ifmmode {p_d} \else ${p_d}$ \fi}

\begin{document}

\title{Probing Dust Settling in Proto-planetary Disks with  ALMA }

\author[Boehler et al.]{Y.~Boehler,$^{1,2,3}$ 
 A.~Dutrey,$^{1,2}$
 S.~Guilloteau$^{1,2}$
 and V.~Pi\'etu$^{4}$\\
$^1$ Univ. Bordeaux, LAB, UMR 5804, F-33270, Floirac, France\\
$^2$ CNRS, LAB, UMR 5804, F-33270 Floirac, France\\
$^3$ Centro de Radioastronom\`ia y Astrof\`isica, UNAM, Apartado Postal 3-72, 58089 Morelia, Michoac\`an, Mexico \\
$^4$ IRAM, 300 rue de la Piscine, 38400 Saint Martin d'H\`eres, France.
}

%


\maketitle %

\begin{abstract}
Investigating the dynamical evolution of dust grains in proto-planetary disks
is a key issue to understand how planets should form. We identify under which
conditions dust settling can be constrained by high angular resolution observations at mm
wavelengths, and which observational strategies are suited for such studies.
Exploring a large range of models, we generate synthetic images of disks with different
degrees of dust settling, and simulate high angular resolution ($\sim$ 0.05-0.3'') ALMA
observations of these synthetic disks. The resulting data sets are then analyzed blindly with
homogeneous disk models (where dust and gas are totally mixed) and the derived disk
parameters are used as tracers of the settling factor.
Our dust disks are partially resolved by ALMA and present some specific behaviors on radial
and mainly vertical directions, which can be used to quantify the level of settling. We
find out that an angular resolution better than or equal to $\sim$ 0.1'' (using 2.3 km
baselines at 0.8mm) allows us to constrain the dust scale height and flaring index with
sufficient precision to unambiguously distinguish between settled and non-settled disks, provided
the inclination is close enough to edge-on (i $\geq$ 75$^\circ$).
Ignoring dust settling and assuming hydrostatic equilibrium when analyzing such disks affects
the derived dust temperature and the radial dependency of the dust emissivity index. The surface
density distribution can also be severely biased at the highest inclinations. However, the derived
dust properties remain largely unaffected if the disk scale height is fitted separately.
ALMA has the potential to test some of the dust settling mechanisms, but for real disks,
deviations from ideal geometry (warps, spiral waves) may provide an ultimate limit on the dust 
settling detection.
\end{abstract}

\begin{keywords}
Stars: formation --- stars: circumstellar matter --- ISM: dust
\end{keywords}

\section{Introduction}

Grain growth and dust settling are two key ingredients in the planetary system formation process.
Recent observational evidences suggest that ISM dust grains start to grow
in the early phase of star formation, as soon as dense pre-stellar cores begin to form.
Theory and numerical simulations predict that in Class II proto-planetary disks, the dust orbiting
the Pre-Main-Sequence (PMS) star continues to grow but also very quickly settles along the mid-plane
in typical characteristic time of a few $10^4$ yrs \citep{Dullemond+Dominik_2004,Fromang+Nelson_2009}.
The growth is the first step towards the formation of even larger solid bodies, which ultimately
culminate with planetary embryos. Settling will speed up this process by favouring grain collisions,
firstly by increasing the relative vertical velocities, as settling acts differently in function of
the dust dynamic properties \citep[see e.g.][]{Birnstiel+etal_2010}, and secondly by concentrating dust close to
the midplane. On top of that, a high dust to gas ratio in this area, can affect the gravitational
stability and control the initial step of the formation of planetesimals \citep{Goldreich+Ward_1973}.

Quantifying the dust evolution process is a complex problem since the two physical
processes (grain growth and dust settling)
are simultaneously shaping the disk. The big grains are expected to fall down relatively quickly to
the mid-plane while only small grains, reflecting the stellar light \citep{Burrows+etal_1996,Roddier+etal_1996},
should remain located on the disk surface, at 3-5 gas scale heights.

At a radius of 100 AU from the central star, typical hydrostatic scale heights range between 10-20 AU or $\sim 0.1 ''$
at the distance of the nearest low-mass star forming regions (D$\sim 140$~pc). Therefore, observing settling requires
both the most sensitive and the most resolving astronomical facilities.

Some evidence of dust settling has been obtained from studies using Near-Infrared (NIR) maps obtained	
by the HST \citep{Duchene+etal_2004}, or by the analysis of the Silicate band at 10$\mu$m \citep{Pinte+etal_2008,Dalessio+etal_2001}. 
IR observations only characterize grain growth for small particles with sizes $a \sim 0.1 - 10~\mu$m, as images at wavelength 
$\lambda$ are mostly sensitive to particles of size $a \simeq \lambda/(2 \pi)$.
 Moreover, as the dust opacity in the NIR is still quite large,
the particles we observe are necessarily located high above the disk plane, typically around 3-5 scale heights \citep{Chiang+Goldreich_1997}.

Contrary to IR, the moderate opacity of the mm/submm domain should probe material
throughout the disk structure. The early bolometric observations of envelopes and disks around young stars
\citep{Beckwith+etal_1990} indicated that both the dust absorption coefficient $\kappa_\nu$ and its
spectral index $\beta$ at mm wavelengths have evolved compared to the ISM dust. However, only spatially resolved observations could
alleviate the ambiguity left by the possible contribution of the inner optically thick core.
Furthermore, contamination of the long wavelengths (longer than 4 mm) flux density by free-free emission
can be substantial and should be removed for proper determination of the spectral index \citep{Rodmann+etal_2006}.
Using the VLA (at 7 mm), PdBI and OVRO to probe the dust properties and the dust disk surface density in CQ Tau, \citet{Testi+etal_2003}
concluded that particle have grown up to sizes as large as $\sim 1$~cm. Similar results were obtained on larger samples in $\rho$ Oph (with ATCA)
and in Taurus-Auriga by \citet{Ricci+etal_2010a,Ricci+etal_2010b}. The overall grain growth in proto-planetary disks thus seems
a well establish fact.

More recently,  \citet{Guilloteau+etal_2011} performed a high angular resolution dual frequency study of disks in the Taurus-Auriga region 
with the IRAM array. Apart from disks with inner holes such as LkCa15 \citep{Pietu+etal_2006}, all sources observed with sufficiently high 
angular resolution (0.4-0.8$''$) exhibit steeper brightness gradient at 3 mm than at 1.3 mm. This is the signature of an evolution of the dust spectral index
with radius, with smaller $\beta$ values near the central star.  The inner part of disks, up to 60-80 AU, appears dominated by large particles
leading to a spectral index $\beta$ below 0.5 between $\lambda = 3$ and 1.3\,mm while beyond 100 AU, $\beta$ reaches value consistent with ISM-like grains
(1.7). This constitutes the first observational evidence of radial variations in dust properties, and the characteristic transition radius
between small and large grains is consistent with recent models of dust evolution in disks by \citet{Birnstiel+etal_2010}.

In this paper, we go one step further and study the impact of dust settling on the disk imaging at mm wavelengths, in order
to define adequate observational strategies to constrain this phenomenon with ALMA.
For this purpose, we utilize the code DISKFIT  \citep{Pietu+etal_2007},  which has been upgraded to take into account the
dust settling. The ALMA simulator developed at IRAM \citep{Pety+etal_2002} is then used
to generate realistic ALMA datasets within the wavelength range 0.5 to 3\,mm. Finally, we analyze these synthetic
observations (pseudo-observations) as real data assuming a vertically uniform dust distribution in order to find out robust criteria of
dust settling. We also explore some hidden degeneracies which may bias our estimate of the dust
properties. We then discuss what would be an ideal ALMA observation.

Our dust disk models are described in  Section \ref{sec:mod}. Section \ref{sec:alma} presents the ALMA predictions (pseudo-observations)
and the method of analysis. We then discuss in Section \ref{sec:discuss} the implications of our results.

\section{Model Description}
\label{sec:mod}

\subsection{Disk Model}
\label{sec:mod:disk}

As in \citet{Guilloteau+etal_2011}, we assume a simple parametric disk model.
In Model 1, the gas surface density is a simple power law with a sharp inner and outer radius:
\begin{equation}\label{eq:power}
    \Sigma_g(r) = \Sigma_{0} \left(\frac{r}{R_0}\right)^{-p} ,
\end{equation}
for $R_\mathrm{int} < r < R_\mathrm{out}$.

In Model 2, the density is tapered by an exponential edge:
\begin{equation}\label{eq:edge}
\Sigma_g(r) = \Sigma_0 \left(\frac{r}{R_0}\right)^{-\gamma}  \exp\left(-(r/R_c)^{2-\gamma}\right) .
\end{equation}
Note that Model 1 derives from Model 2 by simply setting $R_c \rightarrow \infty$ and $p=\gamma$ in the above parametrization.
Model 2 is a solution of the self-similar evolution of a viscous disk in which the viscosity is a power law of the radius
(with constant exponent in time $\gamma$) \citep{Linden-Bell+Pringle_1974}.

The kinetic temperature in the disk mid-plane is also assumed to be a power law of the radius:
\begin{equation}\label{eq:temperature}
    T_k(r,z=0) = T_0 (r/R_0)^{-q} .
\end{equation}
We assume that grains and gas are fully thermally coupled, so that the dust temperature $T_\mathrm{dust} = T_k$.
We shall further assume that the disk is vertically isothermal, $T_k(r,z) = T_k(r,z=0)$.
Models of dust settling show that most of the dust should mostly settle within one scale-height
\citep{Dullemond+Dominik_2004}, therefore assuming that the dust is isothermal, is
at first order a reasonable assumption. The impact of this assumption will be discussed later.
Under hydrostatic equilibrium, the resulting vertical gas distribution
is a Gaussian
\begin{equation}
\rho(r,z) = \frac{\Sigma(r)}{H_g(r) \sqrt{\pi}} \exp{\left(-\left(\frac{z}{H_g(r)}\right)^2\right)} .
\label{eq:rho-evolz}
\end{equation}
With this definition, the gas scale height $H_g$ is:
\begin{equation}
H_g(r) = \sqrt{\frac{2 r^3 k T_k(r)}{G M_* \mu m_H}}  
\label{eq:height}
\end{equation}
with $k$ and $G$ the Boltzmann and the gravitational constants respectively,
$M_*$ the star mass, $\mu$ the mean molecular weight and $m_H$ the mass of the Hydrogen nuclei.
$H_g$ is also a power law of the radius
\begin{equation}
H_g(r) = H_0 (r/R_0)^h ,
\label{eq:height-power}
\end{equation}
with the exponent $h= 3/2-q/2$. The mean molecular weight $\mu$ is equal to 2.6 in our analysis.

\subsection{Dust Properties}

\subsubsection{Mass and grain size distributions}

Dust settling implies local changes in the dust-to-gas ratio, as well as local variations in the grain size distribution, whose
details depend on the mechanism controlling the dust evolution.
We assume here no radial re-distribution of dust: the dust surface density $\Sigma_d$ follows the gas surface density, and at
any radius the \textit{average} (i.e. vertically integrated) dust-to-gas ratio is equal to the standard ratio:
\begin{equation}
\Sigma_d(r)/\Sigma_g(r)  = \zeta_\mathrm{std} = 1/100
    \label{eq:gas-to-dust}
\end{equation}
Eq.\ref{eq:gas-to-dust} ensures mass conservation independently of settling.
The value of $\zeta_\mathrm{std}$ is only a scaling factor for the total
disk mass in non-settled disks, but also affects settling in some specific models.

We further impose that dust settling does not change the overall dust distribution as a function of grain size, and use
a power law size distribution
\begin{equation}
     \frac{\int\mathrm{d}n(a,r,z) dz}{\mathrm{d}a} = n_0 \left(\frac{a}{a_{0}}\right)^{-\psize}~~~~~~ a_{min} \leq a \leq a_{max} .
     \label{eq:dust-size}
\end{equation}
$n_0$ is the number of grains at the reference size $a_0$, $a_{min}$ and $a_{max}$ are the minimum and maximum radius
of the particles and $\psize$ the exponent of the power law \citep[usually taken from 2.5 to 4, e.g.][]{Ricci+etal_2010a}.
While the vertically integrated grain size distribution is a power law (and a fortiori, the disk averaged
grain size distribution), because of the effect of dust settling, the local grain size $n(a,r,z)$
distribution is no longer a power law of $a$.

\subsubsection{Dust emissivity}

The dust emissivity as a function of frequency depends on the dust size distribution and grain composition.
Once the dust size distribution and grain composition are specified, several
methods can be used to derive the emissivity values. This has to be done with grain sizes varying up to 5 to
6 orders of magnitude.
A serious limitation is our poor knowledge of the grain composition and shape. Moreover, several recent
observations and experiments show that the dust spectral index $\beta$ in the Far IR/mm range depends on the dust temperature
\citep{Pollack+etal_1994, Agladze+etal_1996,Coupeaud+etal_2011}.
The Mie theory is the most popular method \citep[see, e.g.][]{Draine_2006} to predict dust emissivities but remains rigorously
exact for spherical grains only. Other methods, such as the Discrete Dipole Approximation (DDA), are heavier to handle
\citep[see, e.g.][and references therein]{Draine_1988, Draine+Flatau_2012} and still suffer from the dust composition and shape limitations.

 One often uses approximate laws for the dust emissivity in the mm
domain, such as a simple power law prescription $\kappa(\nu) = \kappa_0 (\nu/\nu_0)^\beta$.
 Although in general applicable to the molecular clouds where grains remain small, in disks, this approximation
is only valid over a relatively narrow range of frequencies.
Realistic disk grains can result in emissivity curves which cannot be represented in this way at
mm wavelengths, especially when the largest grain become comparable in size to the wavelength \citep{Natta+etal_2004,Ricci+etal_2010a,
Isella+etal_2009}.  Furthermore, in settled disks, such a representation  would no longer be convenient, as relating the effective
$\kappa_0$ and $\beta$ to the dust settling parameters is a non trivial task. Thus a 2-D (r,z) distribution of the dust emissivity
as a function of wavelength needs to be computed once the settling parameters are specified.

Given the important unknowns in the dust geometry and composition, we have elected to use a parametric method to model the dust
emissivity as a function of grain size and wavelength.

Our approach is based on the fact that the emissivity as a function of frequency displays two asymptotic regimes, the
small wavelengths ($ a \gg \lambda$) where the absorption coefficient is dictated by the geometrical cross section, and the long
wavelengths ($a \ll \lambda$) for which a power law applies. These two regimes are connected by a resonance region near $\lambda = 2 \pi a$.
To study the thermal structure of disks, \citet{Inoue+etal_2009} parameterized the emissivity curves by only retaining the two asymptotic
laws. However, at mm wavelengths, the resonant region can contribute significantly to the emissivity. The detailed
behaviour of this resonant region is not critical, as integration over a size distribution will smooth out any fine structure:
only the width and height matter. We thus elected to parameterize the asymptotic regimes and the width and height of this resonant
region in a simple way. The details are given in Appendix \ref{app:grains}.

To integrate over a given distribution in size, the distribution is sampled on discrete bins. We typically use two (logarithmic) bins
per decade in size, except for the smallest sizes (below 1$\mu$m) where 1 bin per decade is used because these small grains contribute
very little to the emissivity at mm wavelengths (and are also less affected by settling effects). Within each bin, the size distribution
is assumed to remain a power law with the same exponent $\psize$ as the integrated grain size distribution. Our selected functional for
the emissivity $\kappa(\nu,a)$ allows analytic integration over this truncated power law size distribution to derive the mean emissivity
per unit mass.
\begin{figure}
\begin{center}
\includegraphics[angle=0,width=\columnwidth]{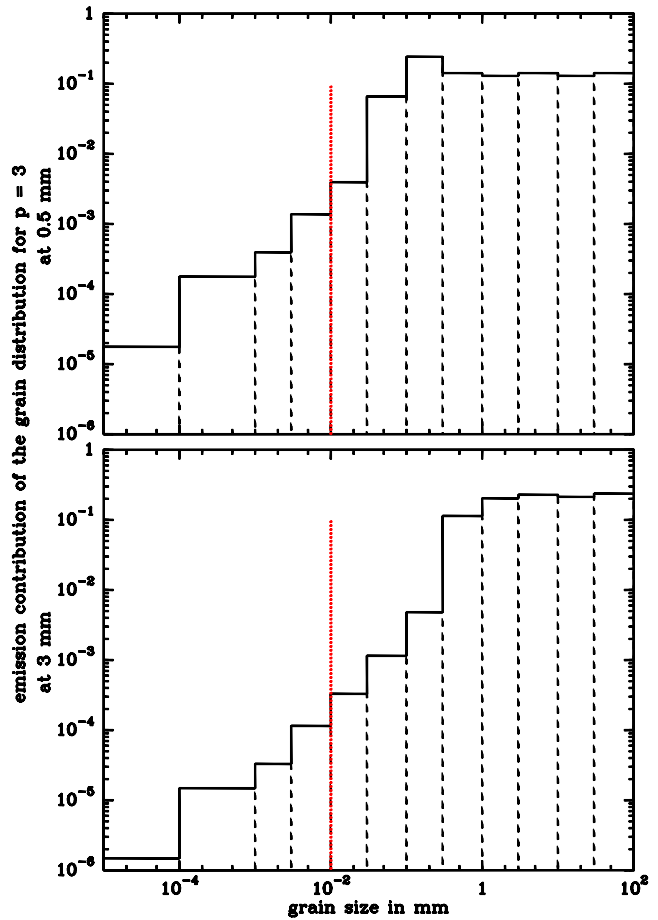}
  \caption{Contribution of various dust grains to the total emission (at 3 and 0.5mm wavelengths), depending of their size, for an assumed
  size exponent $\psize = 3$. The vertical red dotted line represents approximately (depending of disk density) the
  separation between grains well mixed with the gas and grains starting to settle.}
  \label{fig:emission-box}
\end{center}
\end{figure}

In the example presented in this article, the parameters have been adjusted in order to match the dust properties
used by \citet{Ricci+etal_2010a}. The resulting emissivity per size bin are given in Table \ref{tab:emissivity}, and
Fig.\ref{fig:emission-box} shows the relative contribution of each bin to the total emission, for a size
distribution index \psize\  equal to 3.

\begin{table}
\caption{Dust Emissivity calculated from our simplified model.}
\label{tab:emissivity}
\small{
\begin{tabular}{cc|cccc}
\hline
\hline
\multicolumn{2}{c}{grain size} & \multicolumn{4}{c}{$\kappa$ (cm$^{2}$g$^{-1}$)} \\
$a_{-}$ & $a_{+}$ & 0.5 mm & 0.8 mm & 1.3 mm & 3 mm   \\
\hline
0.01$\mu$m    & 30 $\mu$m    & 8.33    & 3.80	 &  1.69   &   0.418	\\
30 $\mu$m    & 100 $\mu$m     & 40.1	& 7.69    &  1.84   &	0.418	 \\
0.1 mm      & 0.3 mm	 & 51.4    & 34.5    &  11.1   &   0.610    \\
0.3 mm     & 1   mm	& 8.60    & 8.74    &  9.52   &   4.12     \\
1  mm	    & 3  mm	 & 2.75    & 2.75    &  2.58   &   2.58     \\
3  mm	    & 10 mm	& 0.860   & 0.860   &  0.826  &   0.826    \\
10 mm	   & 30  mm    & 0.275   & 0.275   &  0.270  &   0.270    \\
30 mm	   & 100 mm    & 0.0860  & 0.0860  &  0.0860 &   0.0859   \\
\hline
\end{tabular}
}
\end{table}

\subsection{Dust Settling}
\label{sec:settling}
Although dust settling mechanism does not in general lead to a Gaussian vertical distribution of grains of a given size $a$, this often
remains an acceptable approximation. Deviations from such vertical profile only occurs high above the typical scale height, i.e. in
regions which contribute very little to the total dust mass \citep[see e.g.][]{Fromang+Nelson_2009}.

It is convenient to define a grain-size dependent scale height, $H_d(a,r)$, and a ``settling factor'', $s(a,r) = H_d(a,r)/H_g(r)$ with
$a$ being the grain size. In our binned dust representation, a \textbf{radius independent} dust settling can be simulated by specifying the values of 
$s_n = s(a_n)$ for each bin. A two-bin representation (one layer of large grains, close to the mid-plane, and one of small, near the disk 
surface) is also used by \citet{Dalessio+etal_2006} to study the impact of dust settling on disk SED. A small difference is that in 
\citet{Dalessio+etal_2006} the two grain categories are spatially separated, while in our case they would only have different scale heights.

To obtain the s(a,r) value, we decided to use instead a more physical approximation based on the results of global numerical
calculations derived from theoretical approaches \citep{Fromang+Nelson_2009} which take into account ideal
MRI-induced MHD turbulence predictions \citep{Balbus+Hawley_1991,Balbus+Hawley_1998} as well as vertical stratification of dust and gas.

In a Keplerian disk, the angular velocity is:
\begin{equation}
\Omega = \Omega_{0} \left(\frac{r}{R_0}\right)^{-3/2}
\end{equation}
and relates to the scale height in hydrostatic equilibrium by:
\begin{equation}
\Omega = \sqrt{2} \frac{C_s}{H_g}
\label{eq:equib-heiht}
\end{equation}
were $C_s$ is the (isothermal) sound speed. The dust stopping time is the typical time, for a particle of size $a$ and density 
$\rho_d$, initially at rest to reach the local gas velocity. In typical T Tauri protoplanetary disks, aerodynamic interactions 
between gas and solid particules smaller than $\sim$ 10 meters are well described by the Epstein regime \citep{Garaud+etal_2004}. We have 
then for the dust stopping time the expression:
\begin{equation}
\tau_s = \frac{\rho_d a}{\rho C_s}
\label{eq:stop_time}
\end{equation}
The main factor controlling the degree of settling is the dimensionless product of the dust stopping time
$\tau_s$ by the angular velocity which fixes the dynamical time. When $\Omega \tau$ $\ll$ 1, the dust particles are
coupled to the gas. When $\Omega \tau$ $\gg$ 1, the dust particles are decoupled from
the gas and settles towards the midplane. This product is linked to the particle size $a$ by:
\begin{equation}
\Omega \tau_s (r,z) = \frac{\sqrt{2 \pi} \rho_d a}{\Sigma_g(r)}  \exp{(z/H_g(r))^2}
\label{eq:link_omtau-a}
\end{equation}
where the surface density $\Sigma_g(r)$ is given by Eqs.\ref{eq:power}-\ref{eq:edge}, depending on
which disk model is used. As this quantity is therefore inversely proportional to the gas surface density,
in general the settling \textit{increases} with radius.

It is convenient to further approximate the effects of dust settling by relating the ``settling factor'' $s(a,r)$
to the settling parameter $\Omega \tau_0 = (\Omega \tau_s)(r,z=0)$:
 \begin{equation}
 s(a,r) = \frac{H_d(a,r)}{H_g(r)} = f \left( \Omega.\tau_0 \right)
 \end{equation}
For large grains, \citet{Dubrulle+etal_1995} and \citet{Carballido+etal_2006} have shown that a power law:
 \begin{equation}
 s(a,r) = \frac{H_d(a,r)}{H_g(r)} =  \left( \Omega.\tau_0 \right)^{\sigma}
 \label{eq:settle_large}
 \end{equation}
with $\sigma = -0.5$ is a suitable function.  With a similar representation, \citet{Pinte+etal_2008} found an
exponent $\sigma = -0.05$ from a multi-wavelength study of IM Lupi. However, their value is mostly constrained
by infrared data, and more specifically the silicate bands which are essentially sensitive to small grains.
We have adopted the following law, which matches the previous asymptotic results
 \begin{eqnarray}
 s(a,r) &=& 1 \hspace{1.75cm} \mathrm{if} \quad{ } \Omega \tau_0 < \omega_c	  \nonumber \\
	&=& \left( \frac{\Omega \tau_0}{\omega_c} \right)^{-0.5} \hspace{0.5cm} \mathrm{if} \quad{ } \Omega \tau_0 > \omega_c
\label{eq:prescription}
 \end{eqnarray}
 where $\omega_c \approx \alpha$, the viscosity parameter,
within  a factor of order unity \citep{Dubrulle+etal_1995}. From Fig.2, we use in red $\omega_c = 6.5\, 10^{-4}$, a
value which slightly overestimates the settling efficiency found by \cite{Fromang+Nelson_2009}. We will also discuss 
in Sec. \ref{sec:omegac} of the value $\omega_c = 1.7\, 10^{-3}$, in blue, which on contrary tends to underestimate it.
For small grains, the small difference between our adopted exponent of 0 for small $\Omega \tau$ and
the value -0.05 found by \cite{Pinte+etal_2008}
is unimportant for our purpose, since the emission in the mm/submm
domain is largely dominated by grains affected by dust settling, as illustrated by Fig.\ref{fig:emission-box}.

\begin{figure}
   \begin{center}
      \includegraphics[angle=270,width=\columnwidth]{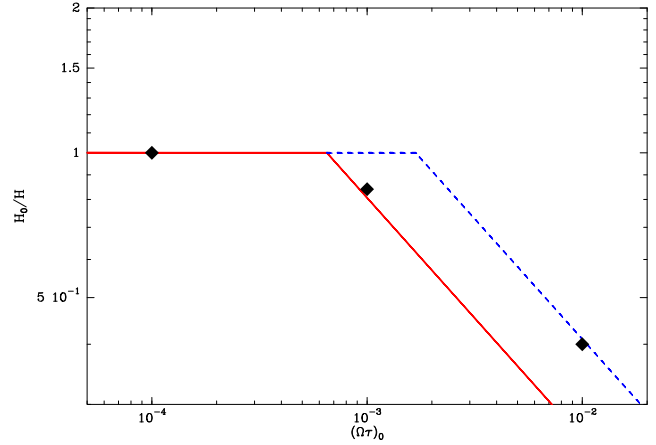} 
      \caption{Dust scale height as a function of $(\Omega.\tau_s)_0$ \citep[adapted from][]{Fromang+Nelson_2009}. 
      The black diamonds represent the values calculated by the simulations.
      The red and dashed blue lines indicate the two functions adopted in our study. }
      \label{fig:from2}
   \end{center}
\end{figure}

\subsection{Radiative Transfer}
We used the ray-tracer of the radiative transfer code DISKFIT \citep{Pietu+etal_2007} to generate brightness distributions
at different wavelengths. As settling can only be observed at sufficiently high disk inclinations, special care was taken
in defining the image sampling to limit the numerical effects, as described in \citet{Guilloteau+etal_2011}.
This required to have radial and vertical cells smaller than  0.05 AU.

\section{Simulations}
\label{sec:alma}


\subsection{Sample of Disk Models}

\begin{table}
\caption{Disk physical parameters for a 1$\Msun$ star.}
\label{tab:pseudo}
\begin{tabular}{c|c}
\hline
\hline
Physical characteristics & Adopted values \\
\hline
type of grains        & Moderate ($\leq$ 3\,mm)  \\
                      & or Large ($\leq$ 10\,cm)  \\
gas scale height      & Hydrostatic Equilibrium (Eq.\ref{eq:height})  \\
Averaged Gas/Dust     & 100 \\
Kinetic Temperature	      & $T_k(r) = 30 \left(\frac{r}{R_0}\right)^{-0.4}$ Kelvin  \\
Dust Temperature	      & $ T_{dust} = T_k $\\
Reference radius      & $R_0$ = 100 AU \\
Inner Radius           & $R_{int}$ = 3 AU \\
Inclinations	      & 70, 80, 85 and 90\textdegree   \\
\hline
\multicolumn{2}{c}{Gas Surface Density (g.cm$^{-2}$)}{Truncated disk (model 1, Eq.\ref{eq:power})}  \\
\hline
$\Sigma_g(r)$       &   $4.35 \left(\frac{r}{R_0}\right)^{-p}$ \\
$p$ & 1 \\
Outer edge	      & $R_{out}$ = 100 AU \\
\hline
\multicolumn{2}{c}{Gas Surface Density (g.cm$^{-2}$)}{Viscous disk (model 2, Eq.\ref{eq:edge})}  \\
\hline
$\Sigma_g(r)$      &  $17.4 \left(\frac{r}{R_0}\right)^{-\gamma}\mathrm{exp}\left(\-(\frac{R}{R_c})^{2-\gamma}\right)$ \\
$\gamma$	      & 0.5 \\
Tapered edge	      & $R_c$ = 50 AU \\
\hline
\end{tabular}
\end{table}

The disks parameters (Table \ref{tab:pseudo}) are representative of the disks studied by
\citet{Guilloteau+etal_2011}.
The disks are in hydrostatic equilibrium with no vertical temperature gradient
and orbit around a $1$\Msun star. The total (gas+dust) disk mass is 0.03 \Msun.

\subsubsection{Dust Settling and Emissivity}

\begin{table}
\caption{Settling factors $s$ for the various grain size distributions}
\label{tab:fn-scale}
\begin{tabular}{cc|cccc}
\hline
\hline
\multicolumn{2}{c}{grain size} & \multicolumn{3}{c}{$s(a) = H_d / H_g$ at Radius (AU) } \\
$a_{-}$ & $a_{+}$ & $R_\mathrm{int}= 3$ & 50  & $R_\mathrm{out} = 100$   \\
\hline
0.01 $\mu$m	 & 10 $\mu$m	 & 1.00    & 1.000   &  1.000	\\
10 $\mu$m      & 30 $\mu$m     & 1.00	 & 0.867   &  0.613   \\
30 $\mu$m      & 100 $\mu$m    & 1.00	 & 0.481   &  0.340   \\
0.1 mm      & 0.3 mm	 & 1.00    & 0.274   &  0.194	\\
0.3 mm      & 1 mm	  & 0.621   & 0.152   &  0.108   \\
1 mm	    & 3 mm	  & 0.354   & 0.0867  &  0.0613  \\
3 mm	    & 10 mm	 & 0.196   & 0.0481  &  0.0340  \\
10 mm	   & 30 mm	& 0.112   & 0.0274  &  0.0194  \\
30 mm	   & 100 mm	 & 0.0621  & 0.0152  &  0.0108  \\
\hline
\end{tabular}\\
The settling factor $s$ is calculated for $p_d=3$, $\rho_d = 1.5$ g.cm$^{-3}$ with $a$
corresponding to the mean (mass weighted) grain radius, and the disk model
described in Table \ref{tab:pseudo}.
\end{table}

\begin{figure}
\begin{center}
\includegraphics[angle=270,width=\columnwidth]{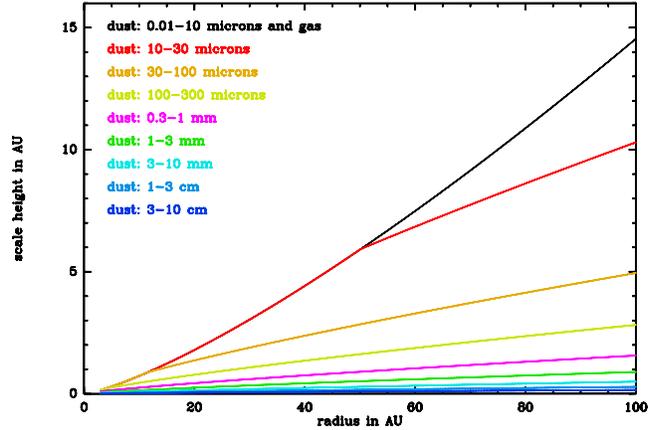}
  \caption{Dust and gas scale heights as a function of the radius for
  different grain sizes for the settled model. The black curve also corresponds to
  the gas scale height.}
  \label{fig:fn-scale}
\end{center}
\end{figure}

We simulate the settling using the prescription of
Eq.\ref{eq:prescription}. Table \ref{tab:fn-scale} gives the corresponding settling factors
and Fig.\ref{fig:fn-scale} indicates the apparent scale
height for various grain sizes as a function of radius.

Following the formalism described in Section 2.2, dust parameters were adjusted to mimic the emissivity curves from
\citet{Ricci+etal_2010a}, see Appendix \ref{app:grains}.
The minimum grain size was 0.01 $\mu$m and the maximum grain size 3\,mm for the moderate grain model or 10\,cm
for the large grain model, with $p_d=3$. We took 9 grain bins for the moderate grains and 12 for the
large ones for ensuring sufficient precision at the ALMA noise level. 
While compact minerals have large specific densities of $\rho_d =  3-4$ g.cm$^{-3}$, we have chosen to use a smaller value
$\rho_d = 1.5 $ g.cm$^{-3}$ to account for the fact that (large) grains are expected to harbor a substantial ice cover
and to be fluffy. The resulting emissivities are given in Table \ref{tab:emissivity}.

\subsubsection{Gas Surface Densities}
The gas surface density used to generate the settled disk model follows either Eq.\ref{eq:power} (power law model, Model 1)
or Eq.\ref{eq:edge} (viscous model, Model 2).
Fig.\ref{fig:from-norm-allf}-\ref{fig:from-norm-allf-prof} were obtained using the Model 1.
Tables 5 and 6 correspond to pseudo-observations using the Model 1. In Table 7 and 8, the pseudo observations
were obtained using the Model 2. The resulting integrated flux densities are given in Table \ref{tab:mod_flux}.

\begin{table}
   \caption{Flux densities (mJy) of settled disks (Model 1)}
   \label{tab:mod_flux}
   \begin{tabular}{lcccc}
     \hline
     \hline
     Frequency  &   100 GHz   &  230 GHz  &   340 GHz	&  670 GHz   \\ \hline
     {Moderate grains} & & & & \\
     70\textdegree & 65  & 450  & 989  & 3490 \\
     80\textdegree & 48  & 300  & 634  & 2100 \\
     85\textdegree & 32  & 180  & 377  & 1240 \\
     90\textdegree &  7.6 & 56  & 137  &  564 \\ \hline
     {Large grains} & & & & \\
     70\textdegree & 9.6  & 60 & 134  & 512  \\
     80\textdegree & 9.0  & 56 & 123  & 462  \\
     85\textdegree & 8.1  & 49 & 107  & 393  \\
     90\textdegree & 2.4  & 17 &  39  & 166  \\
     \hline
   \end{tabular}
\end{table}

\subsubsection{ALMA Configuration}

The simulated brightness distributions obtained from DISKFIT were then processed through
the regularly upgraded ALMA simulator implemented in the GILDAS software package
\citep{Pety+etal_2002} in order to produce the visibilities.

As a first guess, we choose to simulate observations obtained using 50 antennas with
a single antenna configuration, so that observations at different wavelengths can be
performed nearly simultaneously. A maximum baseline length of 2.3 km was used and the
observations were assumed to be around the transit. Pseudo-observations of settled disks,
located at declination $\delta = -23^\circ$, have been created at four different
frequencies, 100, 230, 340 and 670 GHz (or in wavelengths:  3\,mm, 1.3\,mm, 0.88\,mm
and 0.48\,mm, corresponding to the 4 initial ALMA bands 3,6,7 and 9). This leads to a
spatial resolution of 0.30$''$, 0.13$''$, 0.089$''$ and 0.045$''$ for Bands 3,6,7 and
9, respectively. At the distance of the nearest star forming regions (120 -- 140 pc
for $\rho$ Oph and Taurus-Auriga), the corresponding linear resolutions are 39-42, 16-18,
11-12 and 5-6 AU. In our case, we assume a distance of 140~pc. Thermal noise was added
to the simulated $uv$ data (corresponding to 30 min of observations for each frequency).
The resulting image noise (point source sensitivity) are 13 $\mu$Jy at 100 GHz, 20 at 230 GHz,
30 at 340 GHz and 111 at 670 GHz.

Each disk has been imaged at 4 inclination angles ($90^\circ$, $85^\circ$,
$80^\circ$ and $70^\circ$).
The resulting number of visibilities in the pseudo $uv$ tables is 1096704. This number can
be compared to the non reduced $\chi^2$ given in Tables 5 to 8.

\subsection{Prominent Effects of Settling}
To understand the effect of settling, it is useful to compare the images of the same disk (i.e. having the
same gas spatial distribution and mass) with or without settling.
\begin{figure}
  \begin{center}
    \includegraphics[width=\columnwidth]{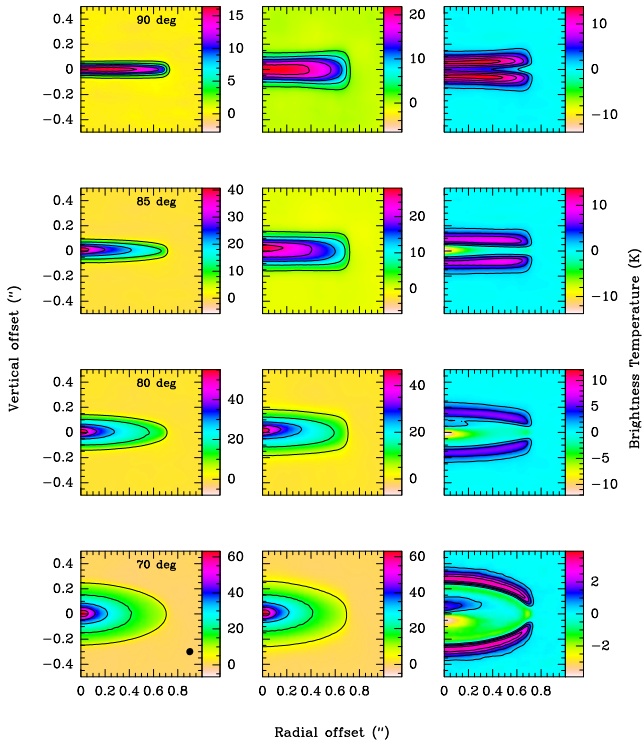}
    \caption{Disks observed at 670 GHz under inclinations of 90$^\circ$, 85$^\circ$, 80$^\circ$ and 70$^\circ$, from
    top to bottom. Left: settled disks. Middle: non-settled disks of same gas mass distribution and same amount of
    dust. Right: difference between these two models (non settled - settled).
    Simulations are made with moderate grains. The hatched ellipse is the PSF.}
    \label{fig:from-norm-allf}
  \end{center}
\end{figure}

\begin{figure}
  \begin{center}
    \includegraphics[angle=0,width=\columnwidth]{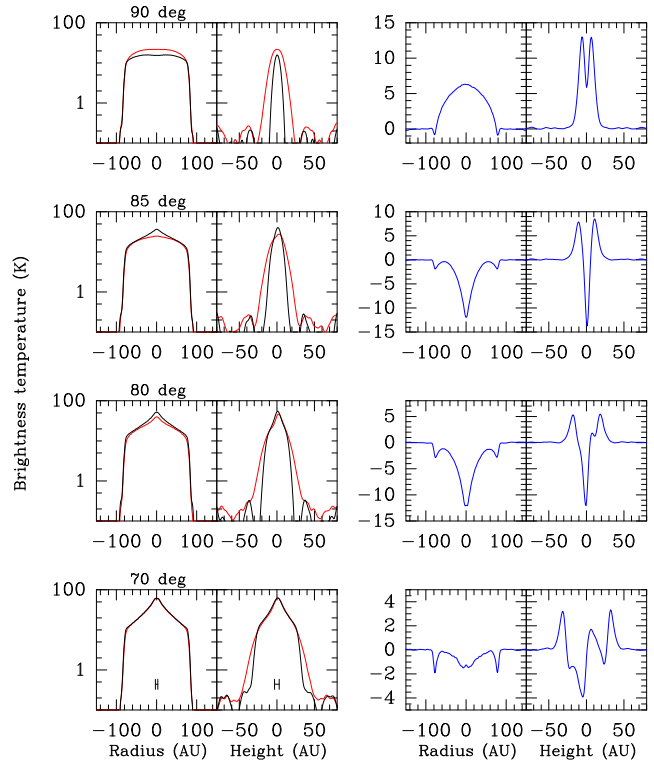}
    \caption{Radial and vertical cuts in brightness temperature distribution (K) of disks with moderate grains,
    observed at 670 GHz under different inclinations. Left: black curves correspond to the settled model.
    Red curves correspond to the non-settled model. Right: the differences (non settled - settled) between
    these two models is shown in blue. The horizontal bar indicates the angular resolution.}
    \label{fig:from-norm-allf-prof}
  \end{center}
\end{figure}

Figure \ref{fig:from-norm-allf} represents the expected images for disks observed at
670 GHz, while Fig. \ref{fig:from-norm-allf-prof} gives brightness profiles for cuts along and perpendicular to the disk midplane at the disk center.

As expected, the vertical extent is smaller in the settled case, as well as the flaring index.
At very high inclinations (only at 90$^\circ$ in our model), the $\tau$ $=$ 1 region of settled disks is reached
at larger radial distances from the star, which are colder. This results in a lower brightness temperature.

We find the same effect with large grains: their lower absorption coefficient is partially compensated by higher column 
densities in the mid-plane due to stronger settling. The self-absorption effect will be smaller for less massive disks. 
Thus, a change in disk mass and a modification of the grain sizes result in different effects, in particular as a function of observing frequency.

Finally, because the intrinsic aspect ratio is of order H/R $\leq$ 0.1, these opacity effects are
critically dependent on the inclination (see Figs \ref{fig:from-norm-allf} and \ref{fig:from-norm-allf-prof}).
At 70$^\circ$, the impact of dust settling becomes in general difficult to see.

\begin{figure}
  \begin{center}
    \includegraphics[angle=0,width=\columnwidth]{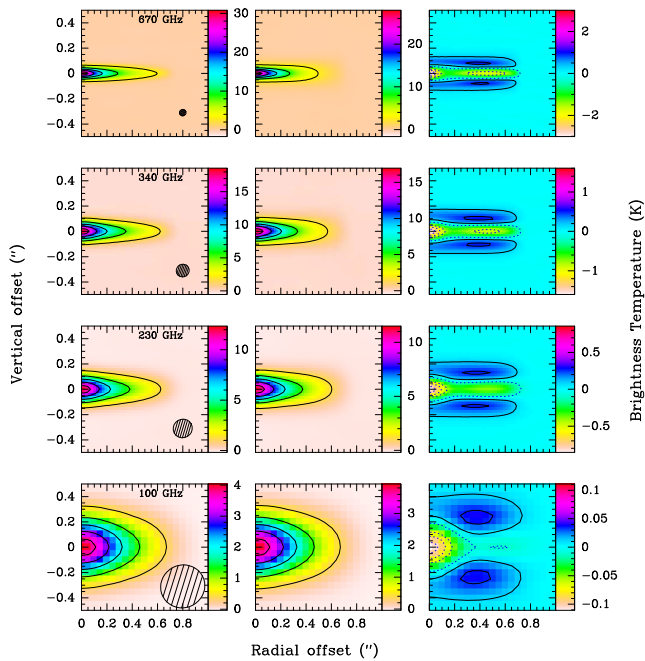}
    \caption{Left: settled disks at 85$^\circ$ of inclination. Middle: results of the best model obtained with
    the non-settled disk model (Case 1). Right: difference between these two models (non settled - settled). Models are
    made with large grains. The hatched ellipse is the PSF.}
    \label{fig:dif-from-ss-gros-85}
  \end{center}
\end{figure}

\begin{figure}
  \begin{center}
    \includegraphics[angle=0,width=\columnwidth]{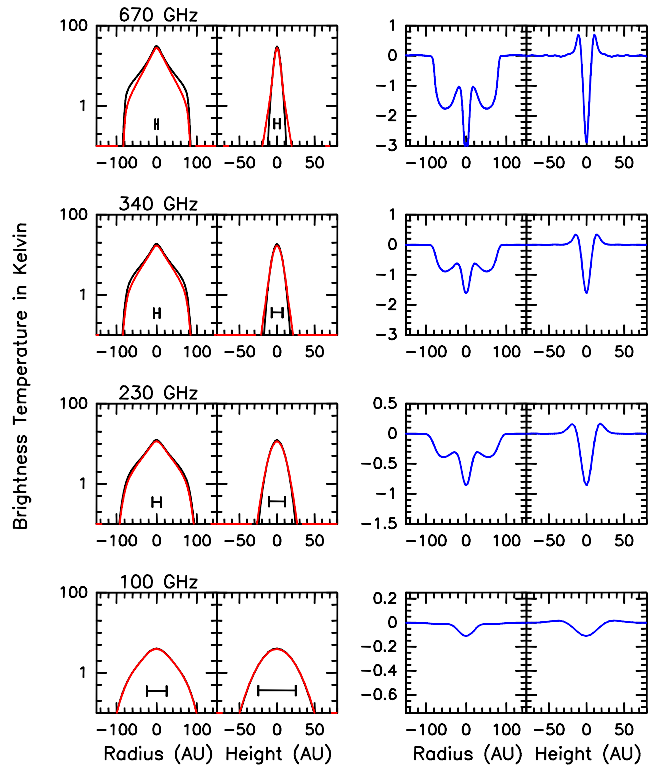}
    \caption{Radial and vertical cuts in brightness temperature distribution (K) at several wavelengths
    for disks inclined at $85^\circ$ and large grains corresponding to Fig.\ref{fig:dif-from-ss-gros-85}.
    Left: black curves correspond to the settled model. Red curves correspond to the non-settled model. Right: the differences
    (non settled - settled) between these two models is shown in blue. The horizontal bar indicates the
    angular resolution.}
    \label{fig:from-ss-gros-85-prof}
  \end{center}
\end{figure}

\subsection{Inversion Process}
\label{sec:inversion}


We investigate here potential ways to distinguish \textit{any} settled disk from \textit{any} non-settled one.
Our approach is to analyze simulated images of settled disks with non-settled, homogeneous disk models.
Under this approach, settled disks may result in very unusual parameters which cannot be ascribed to ``normal'' non-settled
disks. For example, the dust scale height $H_0$ should be small, as well as the flaring index $h$,
in comparison with the hydrostatic scale height.

The resulting $uv$ data sets were fitted by non-settled and vertically isothermal models under the
assumption of power law (Model 1, Eq.\ref{eq:power}) or exponential decay (Model 2, Eq.\ref{eq:edge}) for the surface density distribution.
All frequencies were fitted simultaneously.
Tables 5, 6 and 8 were obtained with minimizations performed using the Model 1 and Table 7 using the Model 1 and Model 2).

\input main-table1.tex
Non-settled disks are characterized by the following parameters: the position angle PA, the inclination $i$, the intrinsic parameters
$R_\mathrm{int}, R_\mathrm{out}, \Sigma_0 ~\mathrm{and~}p$ (for the power law, $R_c, \gamma$ for the viscous model),
$T_0, q$, $H_0, h$ and the dust characteristics. The later being a priori unknown, we assume the
simple power law $\kappa(\nu) = \kappa_0 (\nu/\nu_0)^\beta$ for the dust emissivity.
We use here $\nu_0 = 10^{12}$~Hz and $\kappa_0 = 0.1$~cm$^2$g$^{-1}$ (for a dust to gas
ratio of 1/100). As $\beta$ is a free parameter in our analysis, the choice of $\nu_0$ will affect
$\kappa(\nu)$ at other frequencies, which is compensated in our analysis by adjusting the disk density.
The derived disk density profiles $\Sigma(r)$ (and in particular the disk mass) is thus somewhat
dependent on the assumed value of $\nu_0$.

Each pseudo-observation was fitted with 4 different non-settled disk models.
The scale height was derived either under hydrostatic equilibrium constraint or independently fitted,
and dust emissivity exponent $\beta$ was assumed to be independent of the radius $r$, or evolving like its logarithm:
\begin{equation}
\beta(r) = \beta_i + \beta_r \log (r/R_0)
\label{eq:betadef}
\end{equation}
This leads to 4 cases (see Tables \ref{tab:mod_minimizations}-\ref{tab:big_minimizations}).
Case 1 assumes hydrostatic equilibrium and $\beta_r=0$, Case 2 hydrostatic equilibrium and free $\beta_r$,
while Case 3 uses free scale height $H_0$ and $h$ with $\beta_r=0$ and Case 4 all free parameters
$H_0, h$ and $\beta_r$.
As the impact on $R_\mathrm{int}$ was found to  be non significant in all cases, this
parameter is ignored thereafter.
The disk inclination $i$ is recovered accurately in all cases (with typical error
around 0.2\textdegree),
but its knowledge controls the error bars on some critical parameters, in particular $H_0$ and $h$.
The position angle is also easily recovered, but has less influence than the inclination.

\input main-table2.tex

\section{Discussion}

\label{sec:discuss}

\subsection{Analysis of the Inversion Process}
\label{sec:inversion2}

Tables \ref{tab:mod_minimizations}-\ref{tab:edge_mini_pow}
show the results of the inversion process.
In Tables \ref{tab:mod_minimizations} and \ref{tab:big_minimizations}, both pseudo-observations and
models for the minimizations use the truncated disk surface density (Model 1), with grains of moderate size in
Table \ref{tab:mod_minimizations} and large grains in Table \ref{tab:big_minimizations}. In Table 
\ref{tab:edge_mini}, pseudo-observations, made with the viscous Model 2 and containing large grains, are 
analysed by both Models 1 and 2. 
Finally, Table \ref{tab:edge_mini_pow} refers to pseudo-observations obtained with Model 1 and fitted 
using the Model 2, for moderate size grains.

The case with grains of moderate size illustrates best the problems. It leads to rather strong continuum flux
(Table \ref{tab:mod_flux}), and the optically thick zone is sufficiently large to measure directly the dust
temperature from the surface brightness. The formal errors are very small, indicating that
thermal noise is not a limitation here.

\input main-table3.tex

\subsubsection{Deriving the Scale Height}
When viewed edge-on, the hydrostatic equilibrium assumption (cases 1 and 2) leads to
unusual results. The derived temperature is forced towards low values to better mimic the
small disk thickness ($\sim$ 19 K instead of 30 K).  This is also
true when minimizing a Model 1 by a Model 2 (Table \ref{tab:edge_mini_pow}). A side effect
is an apparent radial dependency of the dust emissivity index ($\beta_r \ne 0$) which is
due to the nonlinearity of the Planck function at low temperatures.
Relaxing the hydrostatic equilibrium
hypothesis (cases 3 and 4) allows us to recover the input temperature profile.

Tables \ref{tab:mod_minimizations}, \ref{tab:big_minimizations} and \ref{tab:edge_mini_pow} also show
that the constraint from the apparent thickness is less important than that from the dust temperature,
so that the fitted scale height in the hydrostatic equilibrium
hypothesis remains unduly large (Fig. \ref{fig:dif-from-ss-gros-85} and Fig.\ref{fig:from-ss-gros-85-prof}).
This artificially creates a deficit of emission close to the mid-plane and an excess at high altitude.\emph{}
 For cases 3 and 4, there is a lack of flaring at all at 90$^\circ$: the settled disk is
 best fitted with a constant thickness. At less extreme inclinations, however, disks appear mildly flared.
 Not only $H_0$ is constrained, but the apparent flaring index $h$ also deviates quite significantly
 in the settled case from the initial value (1.3 in our model, a range between 1.1 -- 1.5 being
 expected for most disks).

\subsubsection{Spectral Index}
Our settled disk are composed of several populations of grains. Each grain population has its own spectral index $\beta$.
If the whole dust emission was optically thin and homogeneously distributed, a mean $\beta$ (defined
as the spectral index between two wavelengths only: 0.5 and 3 mm)
of $\sim 0.61$ for the moderate grains (Table \ref{tab:mod_minimizations}) and $\sim 0.29$ for the large grains (Table
\ref{tab:big_minimizations}) is expected from the opacity curves in Appendix \ref{app:grains}.
The fitted $\beta$ is often different because $\beta$ is not an intrinsic parameter of the dust: our assumed dust properties 
cannot be represented by a single power law between 3 and 0.5 mm, but exhibit
a more complex behaviour (see Appendix). The fitted $\beta$ is more affected for edge-on disks, because the flux densities at
each frequency strongly depend on the degree of settling, thus affecting the relative weights of each observation.

\subsubsection{Degeneracy between $\beta_r$ and $p$}
At very high inclinations (e.g. 90$^\circ$), settling increases the opacity in the disk plane.
A fit of a constant $\beta$ ($\beta_r = 0$) leads to a value of the exponent $p$ of the radial density
profile driven towards negative values, to offer sufficient self-absorption from the cold outer regions.
The independent fit of $H_0$ and its exponent $h$ (case 4) is not sufficient to compensate this effect.
Although this suggests that viscous-like surface density profiles (see Eq.\ref{eq:edge})
with negative $\gamma$ may better fit the images, this is not the case because such profiles drop
too sharply after their critical radius  $R_c$.
Furthermore, there is some ``hidden'' degeneracy between $p$ and $\beta_r$
and the minimization process may converge towards one or the other solution.

\input main-table4.tex

\subsubsection{Impact of the Surface Density Profile}
Table \ref{tab:mod_minimizations} suggests that the scale height can be apparently
constrained independently of the temperature profile even at moderate ($i = 70^\circ$) inclination
(basically all input parameters are recovered properly). This result is due to
the assumed sharp truncation at $R_\mathrm{out} = 100$ AU (Model 1).
The apparent (projected) width of this sharp
edge is a strong indicator of the actual disk thickness.

Table \ref{tab:edge_mini} shows
results for pseudo-observations obtained with a more realistic continuous profile (Model 2,
with $\gamma = 0.5$ and $R_c = 50$ AU). When fitted by a Model 1 (top panel of Table \ref{tab:edge_mini}),
the required scale height is large and the flaring index reaches non physical values of the order of 2.5.
This is an attempt to fit the emission beyond the derived outer radius. On the contrary
when fitted by a Model 2 (bottom panel), a small scale height is indeed recovered. This result indicates
that at inclinations below $80^\circ$, the recovered scale height is sensitive to the exact shape
of the surface density distribution, and cannot in general be determined accurately.
    Table \ref{tab:edge_mini_pow} shows results of tapered disks (Model 2) fitted by
a truncated power law for different inclinations (Model 1) and moderate size grains.
At inclinations $> 80^\circ$, the differences between the true disk density structure and the one assumed in the analysis do
not significantly affect the derivation of the scale height. Other parameters, such as
the temperature, are somewhat affected by the improper surface density profile rather
than by settling.

\subsubsection{Consequences}
For the large grains models, the flux densities and the optical depths are lower. The same
trends are found. Large grains settle more efficiently and the fitted scale height
is even smaller than in the previous case. Since the optically thick core is small,
some degeneracies start appearing between $T,q$ and $\Sigma,p$, as a purely
optically thin emission only depends on $\Sigma T$ and $p+q$.

In all cases, the inconsistencies appearing when fitting by a standard,
non-settled disk model, clearly flag the ``observed'' disk as being unusual, and combined
with the low absolute values of the scale height ($\simeq 2-3$ AU), point towards dust settling
as the only reasonable cause of the discrepancies. Moreover, settled disks actually appear
``pinched'' ($h < 1$)~ rather than flared ($h > 1$). The above analysis
also demonstrates that radius dependent settling as derived from MRI simulations and theoretical
analysis can be distinguished from \textbf{radius independent one}, the later
would not affect the flaring index value $h$. However, directly retrieving the settling factor $s(a,r)$ will
remain largely model dependent, as this would imply to deconvolve from the grain size distribution $n(a)$,
which remains unknown. Even with prior knowledge of $n(a)$, the strong smoothing resulting from this size
distribution would severely limit the capability to retrieve $s(a,r)$ from $H(r)$.

Others parameters like $\beta$, $\beta_r$ or $p$ are sensitive to the dust settling at
inclinations $\geq$ 80$^\circ$ but can only serve as secondary
indicators. In real data, the $\beta_r$ which deviates from its original
value 0 at inclinations $>$ 80$^\circ$ may be due either to dust
settling or to radial variations of the grain properties
\citep{Guilloteau+etal_2011}.

An inclination close to 90$^\circ$ is clearly the more suitable case to study settling since opacity and
brightness temperature effects are maximum. Taken into account the various uncertainties,
in particular on the surface density and radial grain properties, our results suggest
that observations of settling would be possible at inclinations $> 75-80^\circ$.

\subsection{Impact of the various Wavelengths}

\begin{figure}
\begin{center}
\includegraphics[width=6.0cm,angle=0.0]{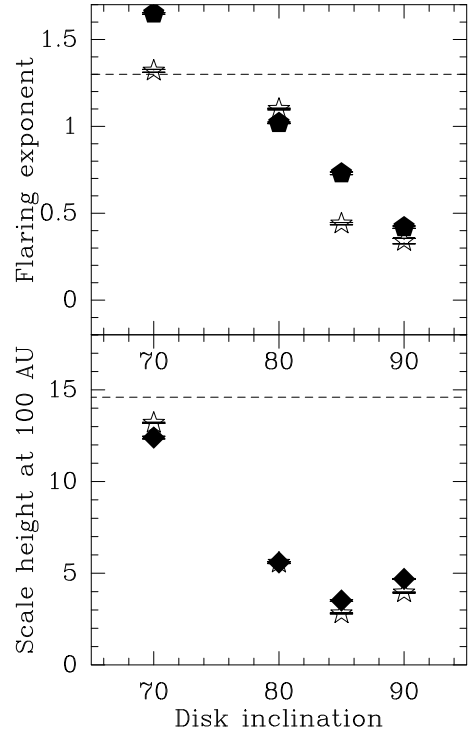}
  \caption{Derived apparent scale heights and flaring index as a function of disk inclination. All results 
  show viscous pseudo-observations (Model 2) fitted by truncated disks (Model 1), using grains of moderate size, 
  with free scale height and radial dependent $\beta$ (case 4). Stars show results obtained with the strong 
  dust settling prescription while filled symbol are obtained with the lowest one.  
  The dashed lines indicate the expected hydrostatic scale height and flaring index for non-settled disks.}
  \label{fig:main_result}
\end{center}
\end{figure}

The above studies show that all the impact of dust settling is only in the effective scale height
(Fig.\ref{fig:main_result})
and a priori we may expect that the highest frequency data, which has the highest spatial resolution,
may be sufficient in itself. This must be moderated by a number of caveats, however.
First, the best signal to noise depends on the dust properties and is not necessarily at the highest frequency. 
Second, the apparent (geometrically constrained) scale height must be compared
to the hydrostatic scale height to prove settling. This implies that a) the (gas) temperature or the
dust temperature as a proxy, should be known, and b) the stellar mass must also be constrained to a reasonable accuracy
(to derive H$_g$, see Eq.\ref{eq:height}). In principle, the gas temperature can be retrieved by imaging thermalized
lines. However, as most chemical models predict that simple molecules lies in a layer about 1-2 scale height
above the disk plane \citep[because of depletion on dust grains in the cold denser regions, see e.g.][]{Semenov+Wiebe_2011},
finding a suitable probe for the disk plane is not straightforward.
In our approach, the dust temperature is derived by resolving
the optically thick parts of the disk. With radial gradients of the dust emissivity index like found by \citet{Guilloteau+etal_2011}
and predicted by simulations of \citet{Birnstiel+etal_2010}, the proper identification of an optically thick core
region requires at least 3 frequencies.
Thus unless some gas temperature can be derived independently, a 3-wavelength study seems required to avoid
ambiguities in identifying dust settling.

The relative ability of each of our 4 observing wavelengths can be evaluated. For the two shortest
ones (0.5 and 0.8\,mm), the errors on the derived parameters (e.g. $H_0$ and $h$) approximately
scale as the wavelength. Since the signal-to-noise ratio is similar at both frequencies, the driving factor
is the angular resolution. The errors then strongly increases for 1.3\,mm, which no longer
has sufficient resolution, while the 3\,mm data are practically unable to provide any
quantitative constraint. Good observing conditions at 0.8\,mm data being much
more frequent than at 0.5\,mm, this wavelength may be the best compromise
in term of sensitivity and angular resolution if only one wavelength can be observed.

We note that the error on $T_0$ in the combined analysis is lower than the simple
weighted average of the 4 independent determinations which shows the
gain in the multi-wavelength approach.

We finally made a last check at 3\,mm on long baselines by using the Model 1 to produce
pseudo-observations in the case of the moderate grain size distribution and assuming an inclination
angle of 85$^\circ$. Baseline lengths of $\sim$ 11 km provide an angular resolution of about 0.06$''$ or 8 AU,
similar to that reached at 0.5\,mm with the $\sim 2$ km baselines. We mimic 4 hours of observations.
We analyzed the pseudo-observations using the non settled disk Model 1 and found that the
scale height is marginally constrained with $H_0 = 2.2 \pm 0.7$ AU and $h= 1.16 \pm 0.3$.
This also barely differs (by $\sim 1.5 \sigma$) from the 4 wavelength fit where we obtain
$H_0 = 3.13 \pm 0.03$ AU. The large errors at 3\,mm  are due to insufficient sensitivity.
Thus, measuring the differential settling between 0.5 and 3\,mm would be very time consuming.

\subsection{Comparison with other imaging simulations}

Using the code MC3D, \citet{Sauter+Wolf_2011} have investigated dust settling by producing
intensity maps of dust disks from 1.0~$\mu$m up to 1.3\,mm. Their analysis differs from ours in
three major points.

First, they only assume two dust grain distributions (small and large) following
the parametrization  proposed by \citet{Dalessio+etal_2006}.
Their small grain population is ISM-like, the large grain distribution extends
up to $a_{max}$ = 1\,mm. This parametrization is similar to the 2-bin version of our \textbf{radius independent}
settling models (Section \ref{sec:mod}). It is very well suited to study
settling in the NIR and Mid-IR because of the high dust opacity but has a too small number
of bins to properly mimic dust settling at mm wavelengths. The maximum grain size
may not be sufficient, as shown for example in Fig.\ref{fig:fn-scale} where the larger
grains significantly contribute to the mm emissions. They also only use stronger settling
parameters, with their large grain scale height smaller than the small grain one by factors
8, 10 or 12. This roughly corresponds to the settling factor in our large grain case,
but the ratio is of the order of 3 for our less extreme grain sizes.

Second, they do not take into account the ALMA transfer function. This is adequate
only with sufficient $uv$ coverage, which is not obtained with short integrations
on very long baselines.

Third, and most importantly, they only compare the settled model with the
non-settled disk in four positions. Such a method, optimized for IR data,
does not use all the information contained in the maps or ALMA observations.
Moreover, as the dust opacity is changing with the wavelength, the optimum
positions should vary accordingly.

Given these differences, comparisons are not straightforward. As expected,
we both find weaker flux and reduced flaring for highly inclined settled disks,
but our method appears much more discriminant and applicable to a wider range
of disk inclinations.

\subsection{Critical discussion}


\subsubsection{Temperature Structure:}
We assume that the temperature is vertically isothermal (as did \citet{Fromang+Nelson_2009}).
In real disks,
the temperature is expected to rise two or three scale heights above the disk plane.
Dust settling will affect this temperature gradient which is mostly
driven by the distribution of small to mid-size grains ($\sim$ 0.1 to 10 $\mu$m)
because they control the opacity to incoming radiation. These grains
exhibit only limited settling.
Indeed, because the apparent scale height at mm wavelengths is a factor 3 to 4
times smaller than the hydrostatic scale height, more than 99 \% of the mm flux is built
in within one hydrostatic scale height, in which temperature gradients should
be negligible. The location of the super-heated layer changes with dust settling, but
not to the point where it will substantially ($\geq$ 50 \%)
affect the temperature within one pressure scale height.

\citet{Hasegawa+Pudritz_2011}
have recently studied the effect of dust settling on the dust temperature using MC3D \citep{Wolf_2003}.
At 5 AU from the star, they found that the dust temperature near the mid-plane
(within $z \leq 0.3$~ scale-height) is somewhat lower in a settled disk than
in a well-mixed one (see their Fig.4). The super-heated layer appears however hotter (60 K instead of 40 K).
The thickness of the impacted cold layer is around 0.5 AU, much too small even for the longest ALMA baselines.
At the larger radii (50 to 100 AU) investigated in our study, the impact of dust settling on the
temperature structure will be much less significant, because the temperature gradients scale
with the dust opacity ($\propto 1/r$ in typical disks),
as well as with incoming radiation flux ($\propto 1/r^2$).

Vertical temperature gradients are however expected to play a role in the apparent scale height
at optical or NIR wavelengths. Indeed for HH30, \cite{Burrows+etal_1996}
derived a much larger scale height from 2 $\mu$m scattered light using the HST than \citet{Guilloteau+etal_2008} from
IRAM PdBI data: this is more likely a manifestation of temperature gradient than of dust settling.

\subsubsection{Settling Shape and Viscosity Parameter: }
\label{sec:omegac}

We have tested a prescription of the settling which has been derived from MRI driven turbulence simulations from
\citet{Fromang+Nelson_2009}. These simulations span a limited range of $(\Omega \tau_s)_0$, and
Figure \ref{fig:from2} suggests that other settling factors may be used.  We also performed simulations
with $\omega_c$ about twice larger (dashed blue curve on Fig.2), leading to smaller settling but
without any major change in our results as can be seen on Fig.8. The dust scale heights are affected by 
at most 30-40 \%, but are still strongly smaller than the gas ($\sim$ 3 AU instead of 15 at R = 100 AU for the 
grain range size 0.1-1mm) and still easily distinguished from the unsettled case. Furthermore,
 the settling degree is also directed linked to the dust specific density, which is generally
assumed to be between 1 and 3 g.cm$^{-3}$. As our grains have a relatively low dust specific density
(1.5 g.cm$^{-3}$), and then are more coupled to the gas, the dust settling degree we generally used can be considered
as medium. The measurable effects on the apparent flaring index indicate that the settling produced  by MRI
can be distinguished, in some cases, from a radially constant settling. For instance,
in the simulation from \citep{Fromang+Nelson_2009}, there is no dead zone, leading to an
underestimate of the dust settling in the inner disk ($r < 10$~AU).

Having measured $H_d/H_g$, it is tempting to directly quantify the viscosity parameter $\alpha$. When this 
ratio is inferior to 1, the settling efficiency is related to it by
\citep{Dubrulle+etal_1995, Carballido+etal_2006}:
\begin{equation}
s(a,r) = \frac{H_d(a,r)}{H_g(r)} \approx \sqrt{\frac{\alpha}{\Omega.\tau}}.
\end{equation}
However, we do not measure $H_d(a,r)/H_g(r)$ as a function of grain size, but only an ensemble averaged with an
a priori unknown size distribution, and a weighting function depending on the dust emissivity as function
of size and wavelength. Furthermore, $\Omega \tau$ scales as the inverse of the gas surface density,
which varies by factor of a few across the disk radius. These two effects are difficult to separate from
the direct impact of $\alpha$ in the above formula.  Thus, $H_d/H_g$ strongly depends on many parameters and
quantifying $\alpha$ in this way appears impracticable.
This conclusion is unfortunately re-inforced by the large integration times which would
be required to measure differential settling between 0.5 and 3 mm, a minimal step to attempt any correction
from the grain size distribution.

Each grain size bin is assumed to follow a Gaussian shape vertical distribution.
Deviation from Gaussianity are expected, as shown by \citet{Fromang+Nelson_2009}
from their MRI simulations. However, as these deviations occur above two-three scale-heights,
they play a minor role in the mm/submm results, like the temperature profile.

\subsubsection{Disk Size:}
We use a disk outer radius of 100 AU, or a similar characteristic
size for the tapered-edge profile. This is consistent with the sizes
found for disks in the Taurus Auriga or $\rho$ Oph regions
\citep{Guilloteau+etal_2011,Andrews+etal_2009,Andrews+etal_2010}.
The median disk outer radius in \citet{Guilloteau+etal_2011} is 130 AU.

\citet{Guilloteau+etal_2011} also show that large grains are essentially
located in the inner 70 to 100 AU.
The existence of a radial gradient in the grain size distribution will
affect the radial dependency of settling.
The presence of smaller grains beyond 100 AU, as suggested by the observational
results of \citep{Guilloteau+etal_2011}, will increase the apparent scale height
there. These outer regions contribute to less than 10 to 30 \% of the total
flux, so this increased scale height will not mask the settling from the inner
regions. If grain growth is maximal near the star, the settling will become
more important in the inner regions. Thus, the radial gradient
of grain size should increase the expected values of the flaring exponent $h$.
While this reduces one of the signature of settling, the effect can easily
be mitigated by a slightly modified analysis method, as the change in $h$
is correlated with the change in grain size. This is amenable to simple
parametrization, at the expense of one additional parameter.

\subsubsection{Disk Mass:}
We adopted a disk mass of $M_d = 0.03\Msun$ which corresponds to flux densities given in Table \ref{tab:mod_flux}
and are similar to those of disks found in e.g. the Taurus region (a factor 2
larger for the small grain case and a factor 2 smaller for the large grain case).
With enough sensitivity, it is preferable to observe disks with low fluxes for
two reasons: 1) at mm wavelength, low flux densities can be a sign of larger
grains ($a > \lambda$) which settle more efficiently and 2) low flux densities may indicate
lower densities which decrease the coupling between gas and dust.
In the ``large grain'' approximation, the settling factor $s$ scales as $\sqrt{M_d/a}$ (from Eq.\ref{eq:stop_time} and
\ref{eq:settle_large}), while for optically thin emission, the flux density $S_\nu$ scales as $\kappa(a) M_d \approx M_d/a$,
so that $s \propto \sqrt{S_\nu}$, the settling efficiency $1/s$ may actually be larger for disks with
lower observed flux densities, unless these low flux values are just due to  disks of similar
intrinsic densities, but smaller outer radii.

\subsubsection{Disk Radial Structure: }
We performed a series of tests where we fit the settled model (using moderate and large grain size distributions)
with a viscous law for the surface density (Model 2) by a non-settled model assuming a power law surface density (Model 1).
The results of the minimization (Table \ref{tab:edge_mini_pow}) show that even if the limited knowledge of the density
profile unavoidably affects the precision with which settling is constrained, it does not mask its existence.
For disks inclined by more than 80$^\circ$, the derived scale height and its flaring
clearly exhibit the behavior expected in case of dust settling. This result is strongly encouraging, especially
as the radial density profile of disks is still a debated issue.

\subsubsection{Disk Geometry: }
 Small departures from perfect geometry and rotational symmetry, like warps and spiral patterns,
 may affect our ability to constrain the scale height. Mis-alignments between jets and disks by a 1-2$^\circ$
 \citep[e.g., HH 30][]{Pety+etal_2006}, and warps of similar magnitude
 \citep[e.g. $\beta$ Pictoris,][]{Mouillet+etal_1997} are known to exists.
They may ultimately limit the apparent scale height to about $H/R \simeq 0.03$, which is comparable
to our ``normal'' grain size case.  The very strong settling predicted for large grains by the MRI turbulent
model may be beyond reach because of this practical limitation.

\begin{figure}
\begin{center}
\includegraphics[width=\columnwidth]{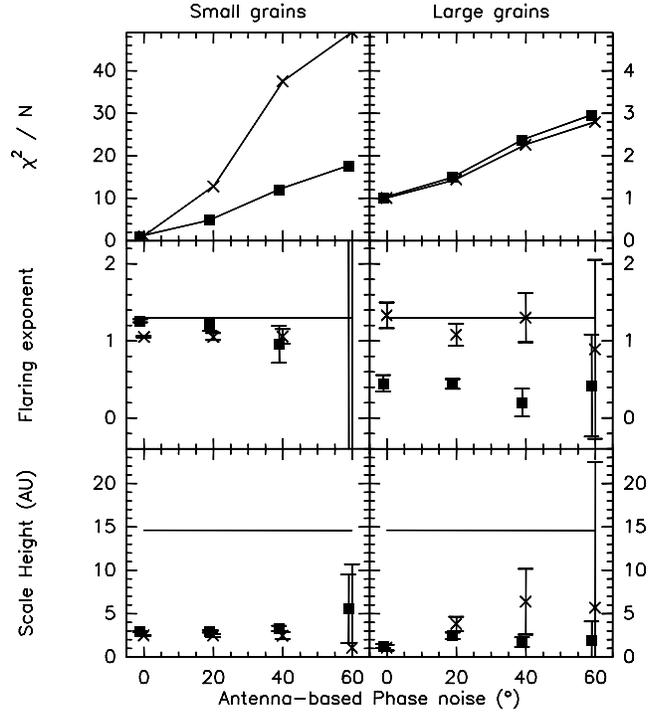}
  \caption{Impact of the phase noise on the measurements of settling. Top: reduced $\chi^2$, middle: flaring
  index $h$, bottom: apparent scale height at 100 AU. Left panels are for the small grain case, and right panels for the big grain case.
  The truncated power law (model 1) is used. Crosses are for $i=70$\textdegree, and squares for $i=85$\textdegree. The horizontal 
  lines indicate the expected values for un-settled disks, derived from the hydrostatic conditions.}
  \label{fig:phase-noise}
\end{center}
\end{figure}

\subsubsection{Instrumental Effects: }

We have shown that thermal noise is not a major limitation
to measure dust settling. We investigated here the impact of
atmospheric phase noise by adding antenna based, Gaussian
distributed phase errors, as would be expected after radiometric
phase correction. Figure \ref{fig:phase-noise} shows the impact on the results.
As expected, the impact is much worse on the brightest sources
(the small grain case). However, for reasonable observing
conditions (antenna based rms noise below 30$^\circ$), phase noise does not prevent the measurement
of the scale height. The flaring index is more affected, but
$h$ still shows significant deviations at 85$^\circ$ of inclination.

\section{Summary}
We have studied how ALMA can be used to quantify the degree of dust settling
in proto-planetary disks around T Tauri stars. We simulated settled disks using
prescriptions for dust settling based on MRI driven viscosity. Using a parametric model to
fit the predicted dust emission as a function of wavelength, we show to what extent settling
can be constrained. Our main findings are
\begin{itemize}
\item For the characteristic dust disk sizes found by previous mm surveys,
dust settling can be measured in typical disks with moderate integration times (of about
2 hours per source), using baselines of the order of 2 km at the
distance of the nearest star forming regions (120 - 140 pc).
\item This is possible only for disks more inclined than $\sim 75 - 80^\circ$.
\item Unless the gas scale height can be independently derived, at least 3 frequencies
are needed to unambiguously identify settling, by comparing the apparent scale height
to the derived dust temperature.
\item The 3\,mm band, which is useful to constrain $\beta$, is less sensitive to settling
than shorter wavelengths even on long baselines (11 km) and for longer integration times
(4 hours). Thus, measuring the differential settling between 0.5 and 3\,mm would be very time consuming.
\item Phase noise should be below about 40$^\circ$ to avoid smearing by limited seeing.
Although the highest frequencies provide better angular resolution, this condition favors the
0.8\,mm band as the preferred frequency to probe the apparent scale height.
\item At the highest inclination ($> 85^\circ$), the apparent radial dependency of the surface density is
affected by dust settling. However, this effect is not a sufficient diagnostic.
\item Other parameters, such as the radial dependency of the dust emissivity index, are
not substantially altered by settling.
\end{itemize}
Our study was performed using a viscosity parameter $\alpha \sim 10^{-3}$. Although settling
is expected to depend on $\alpha$, the dependency is weak ($\sqrt{\alpha}$), and other
unknowns, in particular the grain size distribution but also the surface density,
preclude an accurate determination of $\alpha$ based on the observation of settling only.

\vspace{0.4cm}
\textbf{Acknowledgements:}\\
We acknowledge Sebastien Fromang for useful discussions about
his simulations. The ALMA simulations use the regularly upgraded
ALMA simulator developed at IRAM in the GILDAS package. This research
was supported by the CNRS/INSU programs PCMI, PNP and PNPS and ASA.

\begin{appendix}

\section{Determination of the grain emissivity}

\label{app:grains}

Dust grain emissivities can be computed from their dielectric properties. However, as grain properties are
poorly known in proto-planetary disks, strong assumptions about the grain characteristics (shape, composition, porosity,
ice layer, ...) have to be made for such purpose.

We follow instead a much simpler approach which takes into account the basic asymptotic behaviour of the dust absorption
coefficient as a function of wavelength.
For wavelengths $\lambda$ much smaller than the grain radius $a$, grains behave as optically thick absorbers.  Hence, the
absorption coefficient per unit mass is wavelength independent and
\begin{equation}
\kappa(\lambda,a) = \pi a^2 / m_g = 3 / (4 \rho_d a)
\label{eq:app1}
\end{equation}
for spherical grains of specific density $\rho_d$.
At the other extreme, for $\lambda \gg a$, the emission coefficient usually falls as:
\begin{equation}
\kappa(\lambda,a) \propto  (1/\lambda)^\beta
\label{eq:app2}
\end{equation}
where $\beta$ ranges between 1 or 2, depending on the grain composition but not
on grain sizes.

In between, for $\lambda \approx 2 \pi a$, the absorption coefficient exhibits a number of bumps, due to interferences
between the refracted and diffracted rays. The detailed shape of absorption curve in this resonant region depend
on grain structure \citep{Natta+etal_2004}. However, these  detailed shapes will be smeared
out when the absorption coefficient is computed for a size distribution of the grains, so its exact knowledge is unimportant
provided the overall emissivity curve can be reproduced for realistic grain size distributions.

We thus define the emissivity curve through a small number of parameters.
For a given grain radius $a$, the short wavelength regime is given by:
\begin{equation}
k_s(a) = \frac{3}{4 \rho_d a}
\label{eq:app3}
\end{equation}
from equation \ref{eq:app1}.
The long wavelength regime  is defined by $k_l$ and $e_l$, so that for $\lambda \gg a$,
\begin{equation}
 \kappa(\lambda) = k_l \lambda^{e_l}
\label{eq:app4}
\end{equation}
The two regimes intersect at
\begin{equation}
\lambda_0(a) = a \left( \frac{k_s(a)}{k_l}\right)^{1/e_l}
\label{eq:app5}
\end{equation}

\begin{figure}
   \begin{center}
      \includegraphics[width=\columnwidth]{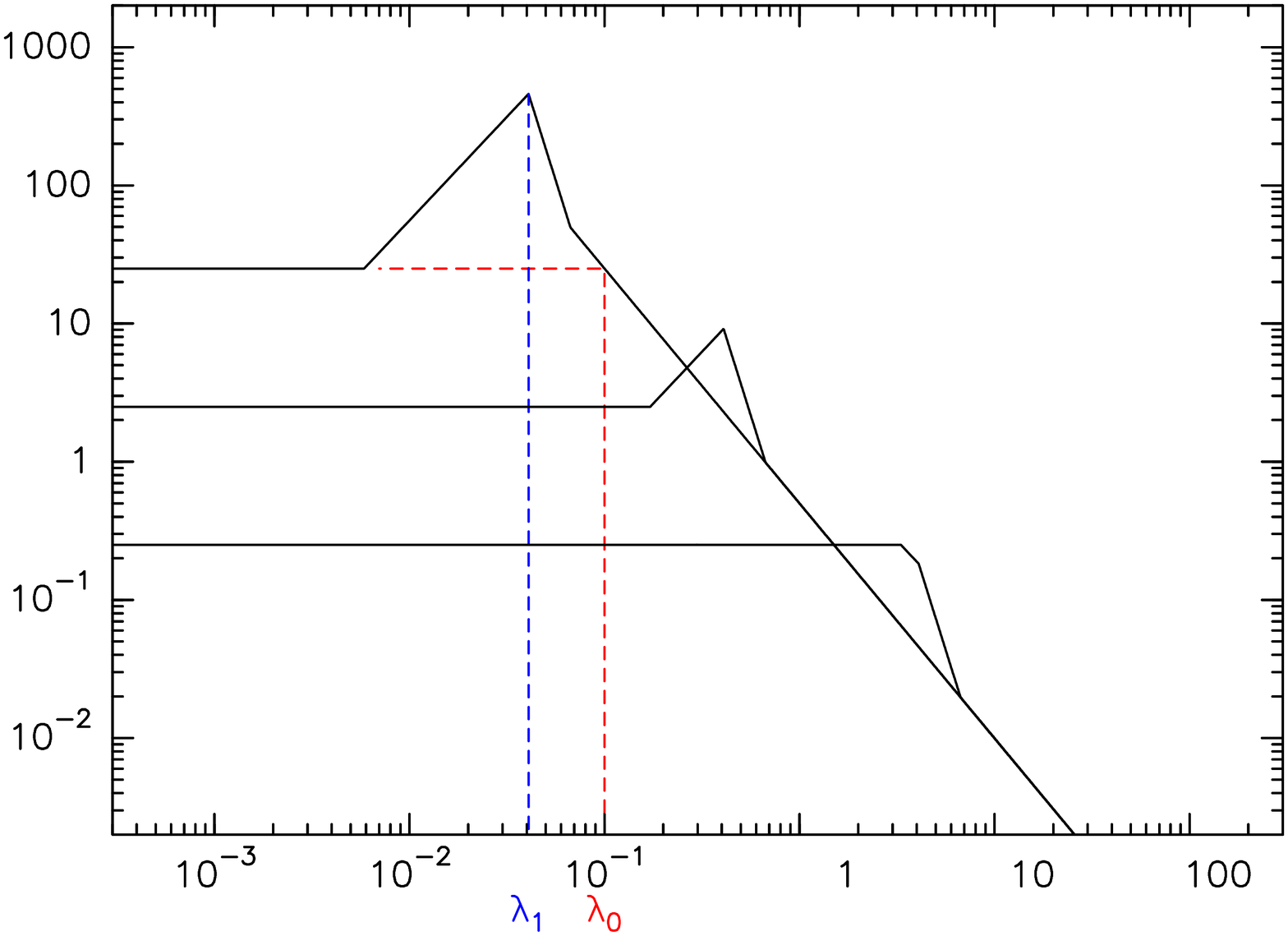}
      \caption{Absorption coefficient as a function $\lambda$ for three different size of grains(0.1, 1 and 10 mm).
      These curves correspond to $\rho_d = 1.5$ g.cm$^{-3}$,  long wavelength
      parameters exponent $e_l = -1.7$ and coefficient $k_l= 0.5$ cm$^{0.3}$.g$^{-1}$ , and bump height $f_p = 4$, 
      exponents $e_b=1.5$, $e_a=-4.5$ and 98	position $l_1 = 0.65$.}
      \label{fig:opa-lambda}
   \end{center}
\end{figure}

The enhanced emissivity (``\textit{bump}'') is defined at $\lambda_1(a) =  2 \pi l_1 a$ by an enhancement factor
$f_p > 1$ compared to the long wavelength asymptotic regime
\begin{equation}
 \kappa(\lambda_1(a),a) = f_p k_l \left(\frac{\lambda_1(a)}{a}\right)^{e_l} =
  f_p k_l \left(2 \pi l_1\right)^{e_l}
\label{eq:app6}
\end{equation}
The shape around this region is defined by slopes $\pm e_b$ before ($\lambda < \lambda_1(a)$)
and $e_a$ after ($\lambda > \lambda_1(a)$) the bump. The $\pm$ sign for $e_b$ occurs because
in this parametrization,  $\kappa(\lambda_1(a),a)$  can be smaller than the short wavelength
asymptotic value $k_s(a)$ (see Fig.\ref{fig:opa-lambda}).  The emissivity law being
a piecewise combination of power laws
of $\lambda$ and $a$, integration over a power law size distribution for the grains is straightforward.

\begin{figure}
\begin{center}
\includegraphics[width=6.0cm, angle=270.0]{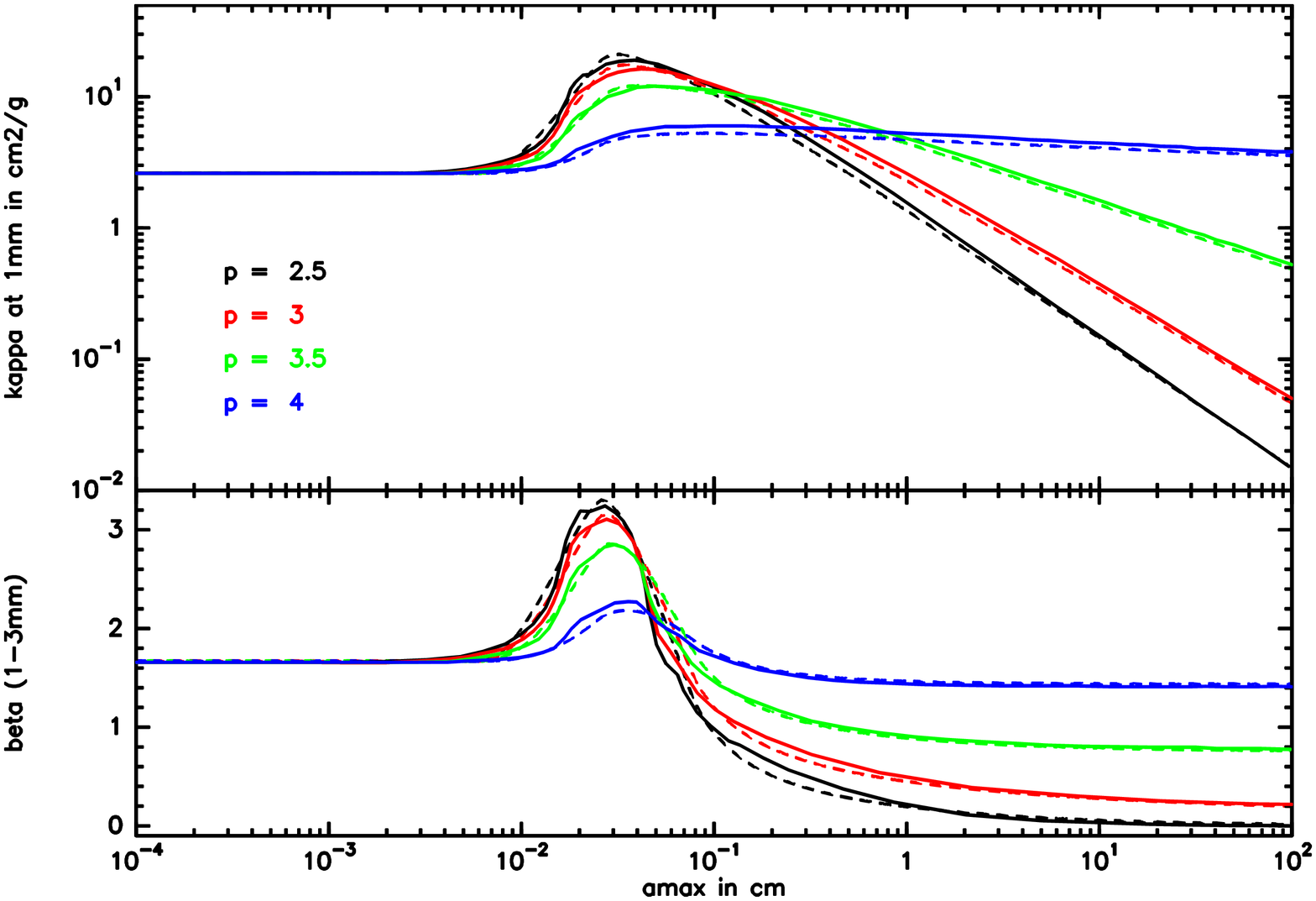}
  \caption{Top panel: Absorption coefficient $\kappa$ at 1 mm as a function of $a_{max}$ for different exponents
  $p$ of the size distribution. The solid lines represent the
  results calculated in Ricci et al (2010), using realistic grains made of astronomical silicates (10\% in volume), carbonaceous
  materials (20\%) and water ice. The dashed lines represent our approximated method.
  Bottom panel: emissivity exponent $\beta$ computed between 1 and 3 mm for the same grain distribution.}
  \label{fig:opa-amax}
\end{center}
\end{figure}

This description with a limited number of parameters captures all the required characteristics
to adequately represent the absorption curves of a given grain size distribution. Figure \ref{fig:opa-amax}
shows the law used in our sample models, compared to the absorption coefficients used by \cite{Ricci+etal_2010a}.
The parameters are $f_p = 15.4$, $e_b = 0.68$, $e_a = -3.5$, $l_1 = 0.65$, $e_l = -1.67$ and
$k_l =  0.058$.
Although differences by 20 \% exist, the key features such as the asymptotic values, position width and height
of the emissivity bump are all well reproduced.

%
%
%
\end{appendix}

\bibliographystyle{mn2e}
\bibliography{biblio}

\end{document}

%% file: main-table1.tex
\begin{table*}
   \caption{Minimizations of a settled disk (model 1) by an homogeneous one (model 1): moderate size grains}
   \tiny
   \begin{tabular}{|l|c|c|c|c|c|c|c|c|c|}
     \hline
     Disk Case \&  & $T_0$           & $q$               & $p$              & $R_\mathrm{out}$ & $\beta_i$   & $\beta_r$                & $H_0$ & $h$ & $\chi^{2}$ \\
       inclination & 30 K & 0.4 & 1.0 & 100 AU & 0.613 & 0  & 14.6 AU & 1.30 &  \\ \hline

     (Case 1) & & & & & & & & &  \\
     70$^\circ$ & 30.3 $\pm$ 0.08 & 0.396 $\pm$ 0.001 & 0.966 $\pm$ 0.002 & 95.8 $\pm$ 0.02 &  0.613 $\pm$ 0.001& [0] & (14.7) & (1.30)  & $1458941 $ \\
     80$^\circ$ & 33.8 $\pm$ 0.07 & 0.378 $\pm$ 0.002 & 1.32 $\pm$ 0.004  & 96.1 $\pm$ 0.03 &  0.420 $\pm$ 0.001& [0] & (15.5) & (1.31) & $1696971 $ \\
     85$^\circ$ & 26.6 $\pm$ 0.10 & 0.629 $\pm$ 0.007 & 0.692 $\pm$ 0.006 & 97.9 $\pm$ 0.04 &  0.341 $\pm$ 0.002& [0] & (13.7) & (1.18) & $1661490 $ \\
     90$^\circ$ & 18.8 $\pm$ 0.22  & 0.810 $\pm$ 0.035 & -1.20 $\pm$ 0.03 & 101.0 $\pm$ 0.07 &  0.892 $\pm$ 0.005& [0] & (11.6) & (1.10) & $1332034 $ \\ \hline

     (Case 2)  & & & & & & & & &  \\
     70$^\circ$  & 30.3 $\pm$ 0.08 & 0.396 $\pm$ 0.001 & 0.965 $\pm$ 0.006 & 95.8 $\pm$ 0.02 & 0.612 $\pm$ 0.002 & 0.000 $\pm$ 0.005 & (14.7) & (1.30) & $1458923 $ \\
     80$^\circ$  & 33.8 $\pm$ 0.13 & 0.378 $\pm$ 0.001 & 1.32 $\pm$ 0.01 &  96.1 $\pm$ 0.03 & 0.420 $\pm$ 0.003 & 0.003 $\pm$ 0.008 & (15.5) & (1.31) & $1697294 $ \\
     85$^\circ$  & 26.4 $\pm$ 0.17 & 0.640 $\pm$ 0.005 & 1.16 $\pm$ 0.01  &  98.4 $\pm$ 0.06 & 0.230 $\pm$ 0.003 & -0.23 $\pm$ 0.01 & (13.7) & (1.18) & $1660374 $ \\
     90$^\circ$  & 19.3 $\pm$ 0.3  & 0.622 $\pm$ 0.013 & 0.55 $\pm$ 0.07  & 102   $\pm$ 0.10 & 0.440 $\pm$ 0.005 & -1.21 $\pm$ 0.02  & (11.7) & (1.19 ) & $1325182 $ \\ \hline

     (Case 3)  & & & & & & & & & \\
     70$^\circ$  & 28.3 $\pm$ 0.04 & 0.423 $\pm$ 0.001 & 1.09 $\pm$ 0.005 & 99.7 $\pm$ 0.007  &  0.686 $\pm$ 0.001& [0] &  2.55 $\pm$ 0.05 & 1.06 $\pm$ 0.016 & 1205325  \\
     80$^\circ$  & 29.2 $\pm$ 0.03 & 0.412 $\pm$ 0.001 & 1.43 $\pm$ 0.009 & 99.7 $\pm$ 0.009  &  0.704 $\pm$ 0.002& [0] &  2.49 $\pm$ 0.04 & 0.94 $\pm$ 0.007 & 1193405  \\
     85$^\circ$  & 30.2 $\pm$ 0.03 & 0.391 $\pm$ 0.001 & 0.27 $\pm$ 0.06 & 100. $\pm$ 0.03 &  0.682 $\pm$ 0.007& [0] & 2.91 $\pm$ 0.06  & 1.18 $\pm$ 0.03  & 1155545  \\
     90$^\circ$  & 31.2 $\pm$ 0.19 & 0.39  $\pm$ 0.034 & -1.20 $\pm$ 0.05 & 101. $\pm$ 0.07 &  0.833 $\pm$ 0.007& [0] & 3.06 $\pm$ 0.05 & -0.09 $\pm$ 0.02 & 1117094  \\ \hline

     (Case 4)  & & & & & & & & & \\
     70$^\circ$  & 28.1 $\pm$ 0.04 & 0.425 $\pm$ 0.001 & 1.22 $\pm$ 0.008 & 99.8 $\pm$ 0.007 & 0.648 $\pm$ 0.002 & -0.080 $\pm$ 0.005 & 2.49 $\pm$ 0.06 & 1.05 $\pm$ 0.014 & 1205325 \\
     80$^\circ$  & 29.1 $\pm$ 0.03 & 0.415 $\pm$ 0.001 & 1.78 $\pm$ 0.02  & 99.8 $\pm$ 0.008 & 0.644 $\pm$ 0.003 & -0.21 $\pm$ 0.02 & 2.52 $\pm$ 0.04 & 0.97 $\pm$ 0.007    & 1192868  \\
     85$^\circ$  & 30.2 $\pm$ 0.03 & 0.391 $\pm$ 0.001 & 0.35 $\pm$ 0.09  & 100 $\pm$ 0.06  & 0.601 $\pm$ 0.007 & -0.46 $\pm$ 0.07 & 2.95 $\pm$ 0.07 & 1.26 $\pm$ 0.03      & 1155401 \\
     90$^\circ$  & 30.2 $\pm$ 0.19 & 0.54  $\pm$ 0.038  & -1.65 $\pm$ 0.09 & 101 $\pm$ 0.07  & 0.897 $\pm$ 0.006  & 0.17 $\pm$ 0.02  & 3.05 $\pm$ 0.06 & -0.11 $\pm$ 0.03 & 1117059 \\ \hline \hline

\end{tabular}\\
    \label{tab:mod_minimizations}
    Numbers between brackets $[]$ indicate fixed parameters. Numbers between parentheses are derived from another parameter ($H_0$ from $T_k$ and $h$ from $q$ under the hydrostatic equilibrium hypothesis). The second row indicates the expected values of the parameters. See section \ref{sec:inversion} for the definition of Cases. 
\end{table*}

%% file: main-table2.tex
\begin{table*}
   \caption{Minimizations of a settled disk (model 1) by an homogeneous one (model 1): large grains}
   \tiny
   \begin{tabular}{|l|c|c|c|c|c|c|c|c|c|}
     \hline
     Disk Case \& &     $T_0$        & $q$               & $p$              & $R_\mathrm{out}$ & $\beta_i$ & $\beta_r$                  & $H_0$ & $h$ & $\chi^{2}$ \\
     inclination & 30 K & 0.4 & 1.0 & 100 AU & 0.288 & 0  & 14.6 AU & 1.30 &  \\ \hline

     (Case 1) & & & & & & & & &  \\
     70$^\circ$ &  22.2 $\pm$ 1.2 & 0.453 $\pm$ 0.014 & 0.92 $\pm$ 0.013 & 99.1 $\pm$ 0.2 &  0.337 $\pm$ 0.004& [0] & (12.6) & (1.27) &  1099461 \\
     80$^\circ$  & 19.2 $\pm$ 0.6 & 0.555 $\pm$ 0.011 & 0.78 $\pm$ 0.01 & 96.5 $\pm$ 0.2 &  0.366 $\pm$ 0.004& [0] & (11.7) & (1.23) & 1103372 \\
     85$^\circ$  & 17.8 $\pm$ 0.3 & 0.668 $\pm$ 0.010 & 0.70 $\pm$ 0.01 & 95.8 $\pm$ 0.2 &  0.464 $\pm$ 0.005& [0] & (11.2) & (1.17) &  1115798 \\
     90$^\circ$  & 11.9 $\pm$ 0.4 & 0.967 $\pm$ 0.055  & -0.82 $\pm$ 0.06 & 96.6 $\pm$ 0.3 &  0.98 $\pm$ 0.02& [0] & (9.1)  & (1.02) & 1113205 \\ \hline

     (Case 2) & & & & & & & & &  \\
     70$^\circ$ &  15.0 $\pm$ 1.0 & 1.10 $\pm$ 0.04 & -0.07 $\pm$ 0.04  & 99.8 $\pm$ 0.1 & 0.45 $\pm$ 0.008 & 0.126 $\pm$ 0.004 & (10.3) & (0.95) & 1099558  \\
     80$^\circ$ &  13.8 $\pm$ 0.9 & 1.30 $\pm$ 0.017 & -0.45 $\pm$ 0.02 & 97.9 $\pm$ 0.2 & 0.50 $\pm$ 0.007 & 0.182 $\pm$ 0.004 & (9.9) & (0.85) & 1102652 \\
     85$^\circ$ &  17.7 $\pm$ 0.24 & 0.674 $\pm$ 0.010 & 0.73 $\pm$ 0.02 & 95.8 $\pm$ 0.2 & 0.45 $\pm$ 0.008 & -0.01 $\pm$ 0.01 & (11.2) & (1.17) & 1114358 \\
     90$^\circ$ &  13.0 $\pm$ 0.35 & 0.660 $\pm$ 0.027 & 0.73 $\pm$ 0.08  & 97.8 $\pm$ 0.2 & 0.43 $\pm$ 0.02  & -0.93 $\pm$ 0.05 & (9.6) & (1.17) & 1112144  \\ \hline

     (Case 3) & & & & & & & & &  \\
     70$^\circ$  & 32.0 $\pm$ 1.0 & 0.40 $\pm$ 0.02 & 1.01 $\pm$ 0.02 & 99.6 $\pm$ 0.06 &  0.266 $\pm$ 0.004& [0] &  1.6 $\pm$ 0.7 & 1.43 $\pm$ 0.13 &  1097745 \\
     80$^\circ$  & 33.0 $\pm$ 0.9 & 0.35 $\pm$ 0.02 & 1.08 $\pm$ 0.02 & 98.9 $\pm$ 0.2 &  0.270 $\pm$ 0.004& [0] &  1.9 $\pm$ 0.5 & 0.68 $\pm$ 0.07 & 1098468 \\
     85$^\circ$  & 33.9 $\pm$ 1.0 & 0.32 $\pm$ 0.02 & 1.13 $\pm$ 0.02 & 98.7 $\pm$ 0.2  &  0.268 $\pm$ 0.004& [0] &  1.4 $\pm$ 0.2 & 0.53 $\pm$ 0.12 & 1099739 \\
     90$^\circ$  & 22.3 $\pm$ 0.8 & 1.49 $\pm$ 0.07 & -1.34 $\pm$ 0.07 & 100. $\pm$ 0.2 &  0.51 $\pm$ 0.02& [0] &  1.4 $\pm$ 0.2 & 0.03 $\pm$ 0.07 & 1096614 \\
     \hline

     (Case 4) & & & & & & & & &  \\
     70$^\circ$  & 27.8 $\pm$ 1.3 & 0.510 $\pm$ 0.05 & 0.84 $\pm$ 0.08 & 99.7  $\pm$ 0.06 & 0.30 $\pm$ 0.01 & 0.027 $\pm$ 0.006 & 1.1 $\pm$ 0.3 & 1.3 $\pm$ 0.4 & 1097740  \\
     80$^\circ$  & 35.9 $\pm$ 1.3 & 0.310 $\pm$ 0.03 & 1.17 $\pm$ 0.03 & 98.9  $\pm$ 0.14 & 0.23 $\pm$ 0.01 & -0.028 $\pm$ 0.005 & 1.6 $\pm$ 0.6 & 0.6 $\pm$ 0.2  & 1098426  \\
     85$^\circ$  & 34.5 $\pm$ 1.1 & 0.309 $\pm$ 0.03 & 1.20 $\pm$ 0.03 & 98.8  $\pm$ 0.2 & 0.22 $\pm$ 0.01 & -0.043 $\pm$ 0.007  & 1.3 $\pm$ 0.2 & 0.45 $\pm$ 0.2  & 1099695  \\
     90$^\circ$  & 58.0 $\pm$ 3.1 & -0.28 $\pm$ 0.03 & 1.64 $\pm$ 0.08 & 100.  $\pm$ 0.2 & 0.11 $\pm$ 0.01 & -0.64 $\pm$ 0.05  & 1.3 $\pm$ 0.2 & 0.05 $\pm$ 0.1  & 1096561 \\ \hline
\end{tabular}\\
    \label{tab:big_minimizations}
\end{table*}

%% file: main-table3.tex
\begin{table*}
   \caption{Tapered Edge disk: large grains}
   \tiny
   \begin{tabular}{|l|c|c|c|c|c|c|c|c|c|}
     \hline
     Disk Case \&  & $T_0$           & $q$               & $p$              & $R_\mathrm{out}$ & $\beta_i$   & $\beta_r$                & $H_0$ & $h$ & $\chi^{2}$ \\
     inclination & 30 K         & 0.4               &                  &                  & 0.288 & 0                       & 14.6 AU & 1.30 &  \\ \hline

     70$^\circ$ & & & & & & & & &  \\
     (Case 1) & 32.9 $\pm$ 0.1 & 0.310 $\pm$ 0.001 & 1.73 $\pm$ 0.002  & 91.7 $\pm$ 0.07 &  0.689 $\pm$ 0.002& [0] & (15.3)  & (1.35) & 1151874 \\
     (Case 2) & 29.8 $\pm$ 0.07 & 0.348 $\pm$ 0.001 & 2.16 $\pm$ 0.004  & 92.9 $\pm$ 0.07 & 0.419 $\pm$ 0.004 & -0.322 $\pm$ 0.004  & (14.6) & (1.33) & 1141912 \\
     (Case 3) & 28.7 $\pm$ 0.11 & 0.379 $\pm$ 0.002 & 1.65 $\pm$ 0.002  & 94.1 $\pm$ 0.08 &  0.725 $\pm$ 0.002& [0] & 20.0 $\pm$ 0.13  & 2.32 $\pm$ 0.007 & 1143237 \\
     (Case 4) & 25.4 $\pm$ 0.07 & 0.426 $\pm$ 0.001 & 1.90 $\pm$ 0.006 & 94.0 $\pm$ 0.08 & 0.629 $\pm$ 0.005 & -0.181 $\pm$ 0.004  & 18.4 $\pm$ 0.14 & 2.31 $\pm$ 0.009 & 1134115 \\ \hline
   \end{tabular}\\
    As Table \ref{tab:mod_minimizations} for a viscous (model 2) settled disk fitted by an homogenous disk with sharp edge (model 1).

   \begin{tabular}{|l|c|c|c|c|c|c|c|c|c|}
     \hline
     Disk Case \&  & $T_0$           & $q$               & $p$              & $R_c$ & $\beta_i$   & $\beta_r$                & $H_0$ & $h$ & $\chi^{2}$ \\
     inclination & 30 AU          & 0.4               & 0.5              &  50 AU                & 0.288 & 0          & 14.6 AU & 1.30 &  \\ \hline

     70$^\circ$ & & & & & & & & &  \\
     (Case 1) & 30.5 $\pm$ 0.09 & 0.387 $\pm$ 0.001 & 0.479 $\pm$ 0.003  & 49.3 $\pm$ 0.04 &  0.620 $\pm$ 0.002& [0] & (14.7) & (1.31) & 1106571 \\
     (Case 2) & 29.6 $\pm$ 0.08 & 0.391 $\pm$ 0.001 & 0.578 $\pm$ 0.003  & 46.2 $\pm$ 0.09 & 0.536 $\pm$ 0.003 & -0.092 $\pm$ 0.002  & (14.5) & (1.30) & 1106488  \\
     (Case 3) & 25.1 $\pm$ 0.10 & 0.473 $\pm$ 0.002 & 0.459 $\pm$ 0.003  & 51.2 $\pm$ 0.05 &  0.682 $\pm$ 0.002& [0] & 3.4 $\pm$ 0.3 & 1.73 $\pm$ 0.03 & 1099314 \\
     (Case 4) & 25.3 $\pm$ 0.09 & 0.463 $\pm$ 0.002 & 0.520 $\pm$ 0.003  & 49.2 $\pm$ 0.09 & 0.628 $\pm$ 0.004 & -0.052 $\pm$ 0.002  & 2.9 $\pm$ 0.4 & 1.65 $\pm$ 0.04 & 1099013 \\ \hline
   \end{tabular} \\
    As Table \ref{tab:mod_minimizations} for a viscous (model 2) settled disk fitted by an homogenous viscous disk (model 2).
    \label{tab:edge_mini}
\end{table*}

%% file: main-table4.tex
\begin{table*}
 \caption{Tapered Edge disk:  moderate size grains}
 \tiny
 \begin{tabular}{|l|c|c|c|c|c|c|c|c|c|}
 \hline
 Disk Case \&  & $T_0$           & $q$               & $p$              & $R_\mathrm{out}$ & $\beta_i$   & $\beta_r$                & $H_0$ & $h$ & $\chi^{2}$ \\
 inclination & 30          & 0.4               &                  &                  &  0.613  & 0                       & 14.6 & 1.30 &  \\ \hline
 (Case 1) & & & & & & & & &  \\
 70$^\circ$ & 34.8 $\pm$ 0.07 & 0.319 $\pm$ 0.001 & 2.14 $\pm$ 0.002  & 103.0 $\pm$ 0.03 &  0.442 $\pm$ 0.001& [0] & (15.7) & (1.34) & 1276606 \\
 80$^\circ$ & 35.0 $\pm$ 0.07 & 0.346 $\pm$ 0.002 & 1.85 $\pm$ 0.004  & 103.0 $\pm$ 0.03 &  0.292 $\pm$ 0.001& [0] & (15.8) & (1.33) & 1415333 \\
 85$^\circ$ & 29.5 $\pm$ 0.13 & 0.504 $\pm$ 0.004 & 1.31 $\pm$ 0.006  & 109.0 $\pm$ 0.06 &  0.293 $\pm$ 0.003& [0] & (14.5) & (1.25) & 1683039 \\
 90$^\circ$ & 19.9 $\pm$ 0.26 & 0.542 $\pm$ 0.008 & 0.16 $\pm$ 0.01   & 109.6 $\pm$ 0.09 &  0.930 $\pm$ 0.006& [0] & (11.9) & (1.23) & 1319208 \\
 \hline
 (Case 2) & & & & & & &  &  &  \\
 70$^\circ$ & 35.5 $\pm$ 0.07 & 0.306 $\pm$ 0.003 & 2.05 $\pm$ 0.005  & 108 $\pm$ 0.03 & 0.484 $\pm$ 0.005 &  0.156 $\pm$ 0.003  & (15.9) & (1.35) & 1299320 ** \\
 80$^\circ$ & 34.8 $\pm$ 0.07 & 0.349 $\pm$ 0.004 & 1.70 $\pm$ 0.007  & 111 $\pm$ 0.03 & 0.483 $\pm$ 0.008 &  0.300 $\pm$ 0.005  & (15.7) & (1.33) & 1393296 \\
 85$^\circ$ & 28.6 $\pm$ 0.09 & 0.533 $\pm$ 0.005 & 1.35 $\pm$ 0.008  & 116 $\pm$ 0.05 & 0.334 $\pm$ 0.010 & -0.016 $\pm$ 0.007  & (14.3) & (1.23) & 1679815 \\
 90$^\circ$ & 20.6 $\pm$ 0.15 & 0.550 $\pm$ 0.009 & 0.18 $\pm$ 0.02   & 111 $\pm$ 0.07 & 0.398 $\pm$ 0.021 & -1.25  $\pm$ 0.013  & (12.1) & (1.22) & 1312764 \\
 \hline
 (Case 3) & & & & & & & & &  \\
 70$^\circ$ & 35.8 $\pm$ 0.1  & 0.306 $\pm$ 0.001 & 2.23 $\pm$ 0.002  & 106 $\pm$ 0.03 & 0.466 $\pm$ 0.001 & [0] & 13.4 $\pm$ 0.04  & 1.33 $\pm$ 0.005 & 1262191 \\
 80$^\circ$ & 35.0 $\pm$ 0.06 & 0.302 $\pm$ 0.002 & 2.26 $\pm$ 0.005  & 110 $\pm$ 0.03 & 0.365 $\pm$ 0.001 & [0] &  6.6 $\pm$ 0.05  & 1.19 $\pm$ 0.006 & 1187642 \\
 85$^\circ$ & 32.8 $\pm$ 0.08 & 0.320 $\pm$ 0.003 & 2.69 $\pm$ 0.009  & 122 $\pm$ 0.05 & 0.461 $\pm$ 0.002 & [0] &  2.8 $\pm$ 0.04  & 0.53 $\pm$ 0.006 & 1129793 \\
 90$^\circ$ & 29.2 $\pm$ 0.13 & 0.294 $\pm$ 0.012 & 0.20 $\pm$ 0.01   & 113 $\pm$ 0.06 & 0.754 $\pm$ 0.076 & [0] &  4.0 $\pm$ 0.04  & 0.10 $\pm$ 0.02 & 1162383 \\
 \hline
 (Case 4) & & & & & & & & &  \\
 70$^\circ$ & 35.9 $\pm$ 0.06 & 0.305 $\pm$ 0.002 & 2.14 $\pm$ 0.004 & 108 $\pm$ 0.04 & 0.493 $\pm$ 0.005 & 0.099 $\pm$ 0.003  & 13.3 $\pm$ 0.03 & 1.32 $\pm$ 0.008 & 1278814 **   \\
 80$^\circ$ & 33.5 $\pm$ 0.08 & 0.330 $\pm$ 0.003 & 2.25 $\pm$ 0.010 & 114 $\pm$ 0.04 & 0.372 $\pm$ 0.005 & 0.172 $\pm$ 0.007  &  5.57 $\pm$ 0.05 & 1.10 $\pm$ 0.005 & 1186973 \\
 85$^\circ$ & 32.4 $\pm$ 0.09 & 0.327 $\pm$ 0.005 & 2.68 $\pm$ 0.017 & 127 $\pm$ 0.06 & 0.451 $\pm$ 0.010 & 0.268 $\pm$ 0.02  & 2.35 $\pm$ 0.03 &  0.44 $\pm$ 0.007 & 1124899 \\
 90$^\circ$ & 34.4 $\pm$ 0.10 & 0.068 $\pm$ 0.004 & 1.21 $\pm$ 0.007 & 122 $\pm$ 0.05 & 0.738 $\pm$ 0.008 & -0.380 $\pm$ 0.005  & 4.04 $\pm$ 0.03 & 0.34 $\pm$ 0.016 & 1145282 \\
 \hline
\end{tabular}\\
    \label{tab:edge_mini_pow}
    As Table \ref{tab:mod_minimizations} for a viscous (model 2) settled disk fitted by an homogenous disk with sharp edge (model 1).
    ** Results probably not converged, as their $\chi2$ is greater than that of the simpler $\beta_r =0$ case.
\end{table*}